\documentclass[a4paper,fleqn,usenatbib]{mnras}

\usepackage{mathptmx}
\usepackage{txfonts}
\usepackage{lscape}

\usepackage[T1]{fontenc}
\usepackage{ae,aecompl}

\usepackage{graphicx}   

\title[BDBS Paper II]{Blanco DECam Bulge Survey (BDBS) II: Project Performance,
Data Analysis, and Early Science Results}
\author[Johnson et al.]{
Christian I. Johnson,$^{1}$\thanks{E-mail: chjohnson1@stsci.edu}
R. Michael Rich,$^{2}$
Michael D. Young,$^{3}$
Iulia T. Simion,$^{4}$ 
\newauthor
William I. Clarkson,$^{5}$
Catherine A. Pilachowski,$^{6}$
Scott Michael,$^{6}$
Andrea Kunder,$^{7}$
\newauthor
Andreas Koch,$^{8}$
A. Katherina Vivas$^{9}$
\\
$^{1}$Space Telescope Science Institute, 3700 San Martin Drive, Baltimore, MD 
21218, USA\\
$^{2}$Department of Physics and Astronomy, UCLA, 430 Portola Plaza,
Box 951547, Los Angeles, CA 90095-1547, USA\\
$^{3}$Indiana University, University Information Technology Services, CIB 2709 
E 10th Street, Bloomington, IN 47401 USA\\
$^{4}$Key Laboratory for Research in Galaxies and Cosmology, Shanghai 
Astronomical Observatory, 80 Nandan Road, Shanghai 200030, China\\
$^{5}$Department of Natural Sciences, University of Michigan-Dearborn, 4901 Evergreen Rd. Dearborn, MI 48128, USA\\
$^{6}$Indiana University Department of Astronomy, SW319, 727 E 3rd Street, Bloomington, IN 47405 USA\\
$^{7}$Saint Martin's University, 5000 Abbey Way SE, Lacey, WA 98503, USA\\
$^{8}$Zentrum f\"ur Astronomie der Universit\"at Heidelberg, Astronomisches 
Rechen-Institut, M\"onchhofstr. 12, 69120 Heidelberg, Germany\\
$^{9}$Cerro Tololo Inter-American Observatory, NSF's National Optical-Infrared Astronomy Research Laboratory, Casilla 603, La Serena, Chile
}

\date{Accepted XXX. Received YYY; in original form ZZZ}

\pubyear{2020}

\begin{document}
\label{firstpage}
\pagerange{\pageref{firstpage}--\pageref{lastpage}}
\maketitle
\begin{abstract}
The Blanco DECam Bulge Survey (BDBS) imaged more than 200 square degrees of
the Southern Galactic bulge using the $ugrizY$ filters of the Dark Energy
Camera, and produced point spread function photometry of approximately
250 million unique sources.  In this paper, we present details regarding the 
construction and collation of survey catalogs, and also discuss the adopted
calibration and dereddening procedures.  Early science results are presented
with a particular emphasis on the bulge metallicity distribution function and
globular clusters.  A key result is the strong correlation ($\sigma$ $\sim$ 0.2 
dex) between ($u-i$)$_{o}$ and [Fe/H] for bulge red clump giants.  We utilized 
this relation to find that interior bulge fields may be well described by
simple closed box enrichment models, but fields exterior to $b$ $\sim$ 
$-$6$\degr$ seem to require a secondary metal-poor component.  Applying scaled
versions of the closed box model to the outer bulge fields is shown to 
significantly reduce the strengths of any additional metal-poor components when
compared to Gaussian mixture models.  Additional results include: a 
confirmation that the $u$-band splits the sub-giant branch in M 22 as a 
function of metallicity, the detection of possible extra-tidal stars along the 
orbits of M 22 and FSR 1758, and additional evidence that NGC 6569 may have a 
small but discrete He spread, as evidenced by red clump luminosity variations 
in the reddest bands.  We do not confirm previous claims that FSR 1758 is part 
of a larger extended structure.

\end{abstract}




\section{Introduction}
The Galactic bulge has been a subject of intense spectroscopic and photometric
investigation over the last 15 years \citep[e.g., see recent 
reviews by][]{Rich13,Babusiaux16,McWilliam16,Nataf17,Barbuy18}.  Numerous lines
of observational and theoretical evidence \citep[e.g.][]{McWilliam10,Nataf10,
Saito11,Wegg13,Ness16,Portail17,Simion17} indicate that the central bulge is 
dominated by a boxy/peanut or ``X"-shape structure.  Kinematic surveys have 
conclusively shown that the majority of bulge stars exhibit cylindrical 
rotation \citep{Howard09,Kunder12,Ness13_kinematic,Zoccali14} consistent with 
properties found in ``pseudobulges" of other galaxies 
\citep[e.g.,][]{Kormendy04}, and in fact the Milky Way may be an almost pure 
disk galaxy \citep{Shen10}.  

However, some metallicity dependent kinematic differences, such as the presence
(or not) of a vertex deviation \citep{Soto07,Babusiaux10} and differing
orbital anisotropies \citep{Clarkson18}, suggest that the Milky Way bar is 
predominantly supported by metal-rich stars.  More metal-poor stars, especially
outside $\sim$1 kpc from the Galactic center, may represent a combination of 
inner halo or thick disk populations \citep[e.g.,][]{Portail17}.  A minor 
classical/merger-built bulge component has not yet been ruled out, and 
chemodynamic evidence from RR Lyrae stars in particular indicates that the 
oldest, most metal-poor bulge stars form either a kinematically hot bar 
\citep{Pietrukowicz12,Pietrukowicz15} or a pressure supported spheroidal 
population \citep{Dekany13,Kunder16,Prudil19}.  The conflicting kinematic 
patterns between stars with different metallicities suggest that the Milky Way
bulge/bar is a composite system

From a chemical standpoint, little agreement has been reached regarding the
bulge/bar's true composition pattern.  Early work indicated that inner bulge
fields near Baade's Window are well-described by closed box one-zone gas 
exhaustion models \citep[e.g.,][]{Rich90,Matteucci90}.  However, subsequent
analyses that included outer bulge fields revealed the presence of a vertical 
metallicity gradient \citep{Zoccali08,Johnson11,Johnson13_offaxis,Gonzalez13},
and many modern large sample surveys claim to find anywhere from 2-5 
populations with distinct [Fe/H]\footnote{[A/B] $\equiv$ 
log(N$_{\rm A}$/N$_{\rm B}$)$_{\rm star}$ $-$
log(N$_{\rm A}$/N$_{\rm B}$)$_{\sun}$ for elements A and B.} values 
\citep[e.g.,][]{Hill11,Ness13_mdf,Bensby17,Rojas17,Zoccali17,GarciaPerez18,
Duong19}.  Although some ``peaks" in the bulge's metallicity distribution 
function may represent contributions from the halo and disk, the true nature of
any metallicity gradient, the actual number of distinct components existing in 
the bulge, and whether a metallicity gradient extends into $\vert$$b$$\vert$ 
$<$ 2$\degr$ \citep[e.g.,][]{Rich12,Schultheis15,Ryde16,Schultheis19} remain 
open questions.

Further issues regarding the formation, evolution, and structure of the 
bulge are raised when considering the detailed chemical abundances and age 
estimates.  The inflection point in plots of [$\alpha$/Fe] versus [Fe/H] for 
the metal-poor bulge ($-$1 $<$ [Fe/H] $<$ 0) and thick disk is a particular 
area of contention.  Some studies find that the bulge remains $\alpha$-enhanced
to a higher metallicity than the local thick disk, which would be consistent
with a more rapid enrichment time scale \citep[e.g.,][]{Zoccali06,Fulbright07,
Johnson11,Bensby13,Johnson14,Bensby17,Rojas17}.  However, others find that the 
bulge and local thick disk trends are identical \citep[e.g.,][]{Melendez08,
AlvesBrito10,Gonzalez11,Jonsson17,Zasowski19}.  Similar disagreements are also
found when considering the light and heavy elements \citep[e.g., see reviews 
by][]{Rich13,McWilliam16,Barbuy18}, and small composition differences may exist
between the inner and outer bulge populations as well 
\citep[e.g.,][]{Johnson12_bulge}.  

Although bulge stars are overwhelmingly old with an age of $\sim$10 Gyr 
\citep[e.g.,][]{Ortolani95,Zoccali03,Clarkson08,Valenti13,Renzini18,Surot19}, 
microlensed dwarf analyses have suggested that a significant fraction of 
metal-rich bulge stars are $<$ 3-8 Gyr old \citep{Bensby11,Bensby13,
Bensby17}.  Furthermore, \citet{Saha19} claim to have found a conspicuous 
young ($\sim$1 Gyr) population of ``blue loop" stars in Baade's 
window.  However, nearly all color-magnitude diagram (CMD) investigations 
have ruled out a young bulge population that exceeds the few per cent level
\citep{Clarkson08,Valenti13,Renzini18,Surot19}, and the bulge's RR Lyrae
population may even contain some of the oldest known stars in the Galaxy
\citep[e.g.,][]{Savino20}.  On the other hand, \citet{Ness14} suggests that 
young bulge stars may be tightly constrained near the plane, and 
\citet{Haywood16} claims that young stars may be difficult to
detect in typical CMDs due to degeneracy between age and metallicity.  A
clear solution to the bulge's age distribution problem is still lacking but 
will be required to fully understand its formation history.

The bulge's accretion history remains an important but unknown quantity as well,
and may be closely connected to the existing observational evidence of a 
classical bulge component.  For example, \citet{Pietrukowicz15}
discovered the existence of two distinct RR Lyrae populations in the bulge that
exhibit separate sequences in period-amplitude diagrams, and \citet{Lee16} 
suggested that the two populations may be driven by He abundance differences
following a pollution mechanism similar to those thought to occur in globular
clusters.  As a result, a significant portion of the Galactic bulge may have 
been accreted from disrupted ``building blocks", such as globular clusters.  
In fact, \citet{Ferraro09,Ferraro16} interpret the peculiar multiple 
populations in Terzan 5 as evidence that this cluster is a surviving example
of a primordial bulge component.  

Further examples of accretion include the presence of possible dwarf nuclei 
cores such as the globular clusters NGC 6273 \citep{Johnson15_6273,
Johnson17_6273} and FSR 1758 \citep{Barba19}, double mode RR Lyrae with 
peculiar period ratios \citep{Soszynski14,Kunder19}, N-rich \citep{Schiavon17} 
and Na-rich \citep{Lee19} stars with potential cluster origins \citep[but see
also][]{Bekki19}, high velocity stars with retrograde orbits 
\citep[e.g.,][]{Hansen16}, and the existence of at least one r-process enhanced
star with a possible dwarf galaxy origin \citep{Johnson13_rII}.  Many of these 
objects have been found in the outer bulge, which suggests that the inner and 
outer bulge may trace different formation and/or enrichment paths.

Several of the results described above were products of spectroscopic surveys 
such as the Bulge Radial Velocity Assay \citep[BRAVA;][]{Rich07,Kunder12}, 
Abundances and Radial velocity Galactic Origins Survey 
\citep[ARGOS;][]{Freeman13}, GIRAFFE Inner Bulge Survey 
\citep[GIBS;][]{Zoccali14}, Gaia-ESO Survey \citep{Gilmore12}, and Apache 
Point Observatory Galactic Evolution Experiment \citep[APOGEE;][]{Majewski17} 
along with photometric surveys such as the Vista Variables in the Via Lactea
\citep[VVV;][]{Minniti10}, Optical Gravitational Lensing Experiment 
\citep[OGLE;][]{Udalski15}, and Two Micron All Sky Survey 
\citep[2MASS;][]{Skrutskie06}.  Due to the large and variable extinction across
most of the bulge, many of these surveys are optimized for near-infrared (IR)
observations.  However, deep optical and near-ultraviolet (UV) observations are
still feasible along most bulge sight lines, and improved reddening maps 
\citep[e.g.,][]{Gonzalez13,Simion17} from near-IR surveys make accurate, large
scale extinction corrections possible.  The inclusion of optical, and 
particularly near-UV, photometry in the bulge significantly enhances the 
impact of previous surveys and permits detailed investigations into the bulge's
metallicity distribution, structure, and UV-bright populations.

Given the paucity of uniform optical and near-UV photometry in the bulge, we
present details regarding project performance, analysis methods, and early
science results from the Blanco DECam Bulge Survey (BDBS).  BDBS takes 
advantage of the Dark Energy Camera's 2.2 degree field-of-view 
\citep{Flaugher15} to survey $>$ 200 contiguous square degrees of the Southern 
Galactic bulge (see Fig.~\ref{fig:density_map}) in the $ugrizY$ filters.  BDBS 
extends to higher Galactic latitudes than the primarily optical DECam Plane 
Survey \citep[DECaPS;][]{Schlafly_2018}, and also reaches lower declinations 
than the Panoramic Survey Telescope and Rapid Response System 
\citep[Pan-STARRS;][]{Chambers_2016} project.  Critically, BDBS includes the
$u$-band, which we show to be an efficient metallicity discriminator for red
clump giants.

\section{Observations and Data Reduction}
\begin{figure*}
\includegraphics[width=\textwidth]{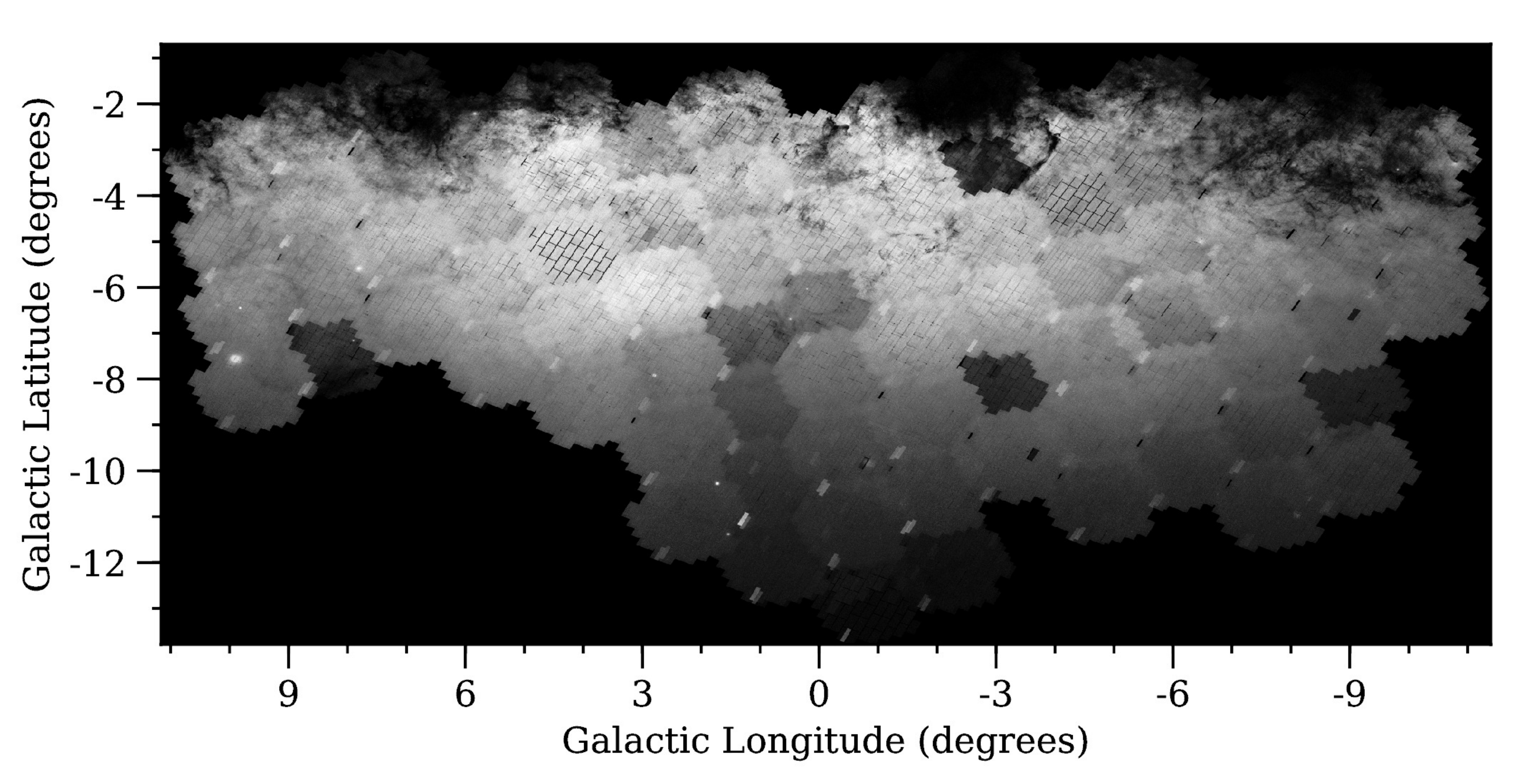}
\caption{A binned source density map comprising 243,959,076 objects is shown
for the contiguous BDBS footprint.  Substantial extinction limits optical
depth along lines-of-sight with $b$ $\ga$ $-$3$\degr$ while incomplete
observations and/or poor observing conditions resulted in fewer detections for
a small number of fields (identified here by lower mean intensity levels).
More than 25 globular clusters are visible, including notable objects such M 22
($l$,$b$) = ($+$9.89,$-$7.55) and FSR 1758 ($l$,$b$) = ($-$10.78,$-$3.29).  The
DECam field-of-view is also visible for some fields where multiple dithers
could not be obtained.}
\label{fig:density_map}
\end{figure*}

The data presented here were obtained using the wide-field DECam imager mounted
on the Blanco 4m Telescope at Cerro Tololo Inter-American Observatory during
windows spanning 2013 June 01-05, 2013 July 14-15, and 2014 July 14-21.  As
illustrated in the $r$-band source density map of Fig.~\ref{fig:density_map}, 
the BDBS footprint provides nearly contiguous coverage from $l$ $\sim$ 
$-$11$\degr$ to $+$11$\degr$ and $b$ $\sim$ $-$2$\degr$ to $-$9$\degr$ , and 
extends down to $b$ $\sim$ $-$13$\degr$ within 3$\degr$ of the minor axis.  
Approximately 90 unique DECam ``pointings" were required to cover the desired 
sky area while also allowing for $\sim$ 10 per cent area overlap between 
adjacent fields.

\begin{figure}
\includegraphics[width=\columnwidth]{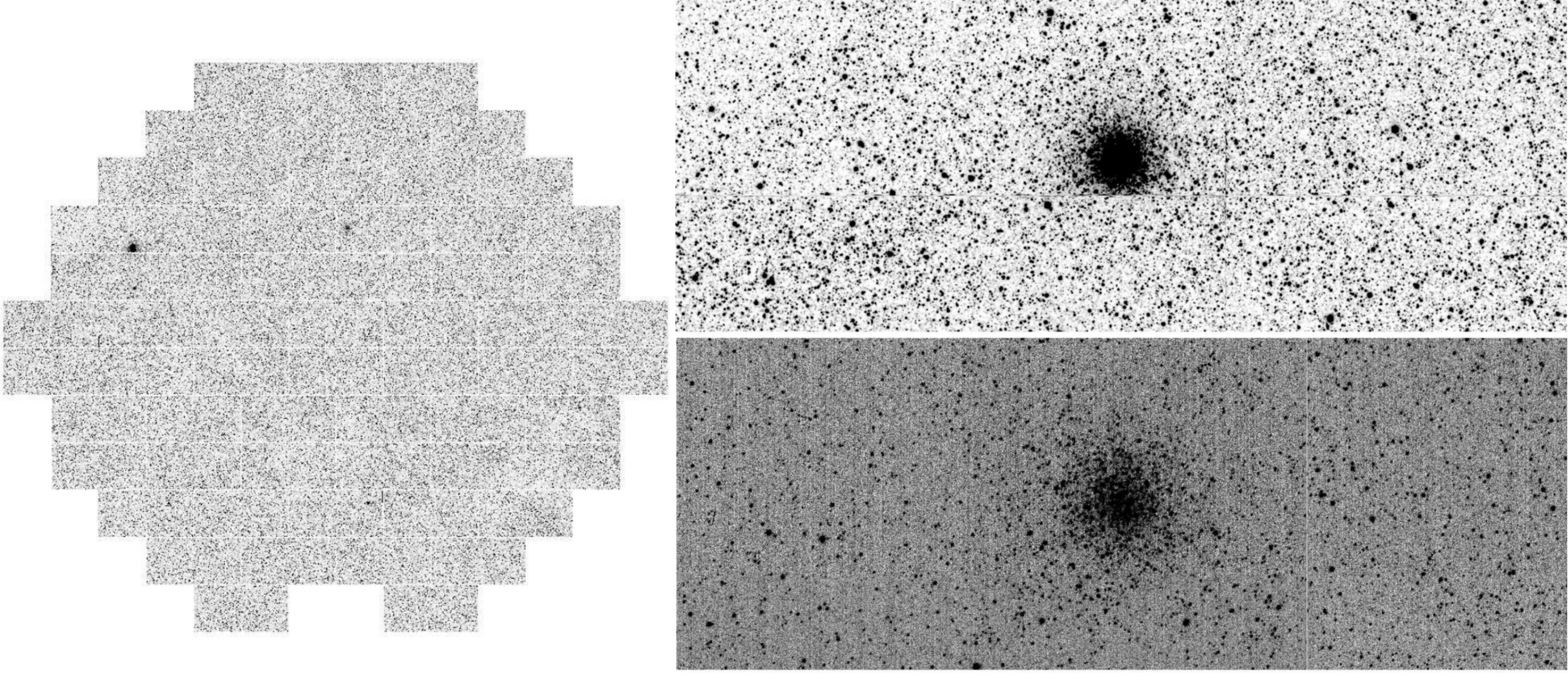}
\caption{\emph{Left:} A sample 75 second DECam $r$-band exposure of a field near
($l$,$b$) $\approx$ ($-$0.21,$-$6.29), which includes the globular clusters
NGC 6569 (left) and NGC 6558 (middle), is shown.  Approximately
1.9$\times$10$^{\rm 6}$ sources were detected in this image.  \emph{Right:} A
zoom-in of the region near NGC 6569 is shown for a ``long" (75 second; top) and
``ultra-short" (0.25 second; bottom) $r$-band exposure.  North is up and East
is to the left for all panels.}
\label{fig:decam_sample}
\end{figure}

For all filters, the ``long" exposures typically followed a four or five point
dither pattern to fill in the substantial gaps between each of the 61 
CCDs\footnote{DECam nominally has 62 CCDs but detector N30 is not used due to
an over-illumination event that occurred in November 2012 (see 
http://www.ctio.noao.edu/noao/content/Status-DECam-CCDs).}; however, for the 
$r$-band an additional, smaller five point dither pattern was used to ensure 
adequate sampling of the point spread function.  These observations are 
intended to serve as a first epoch for future work aimed at obtaining 
DECam-only proper motions.  ``Short" and ``ultra-short" exposures
were also obtained in order to mitigate saturation issues with bright bulge 
red giant branch (RGB) stars, but these observations generally only followed a 
two point dither pattern.  For the $grizY$ bands, the long, short, and 
ultra-short exposures were of order 75, 5, and 0.25 seconds, respectively.  
The long, short, and ultra-short exposure times for the $u$-band were typically
150, 30, and 1.5 seconds, respectively.  A sample DECam field-of-view for the 
$r$-band, along with comparisons of the globular cluster NGC 6569 in long and 
ultra-short exposures, is provided in Fig.~\ref{fig:decam_sample}.  Note that 
all observations were taken with 1$\times$1 binning, which is equivalent to a 
plate scale of $\sim$ 0.263$\arcsec$ per pixel \citep{Flaugher15}.

\begin{figure}
\includegraphics[width=\columnwidth]{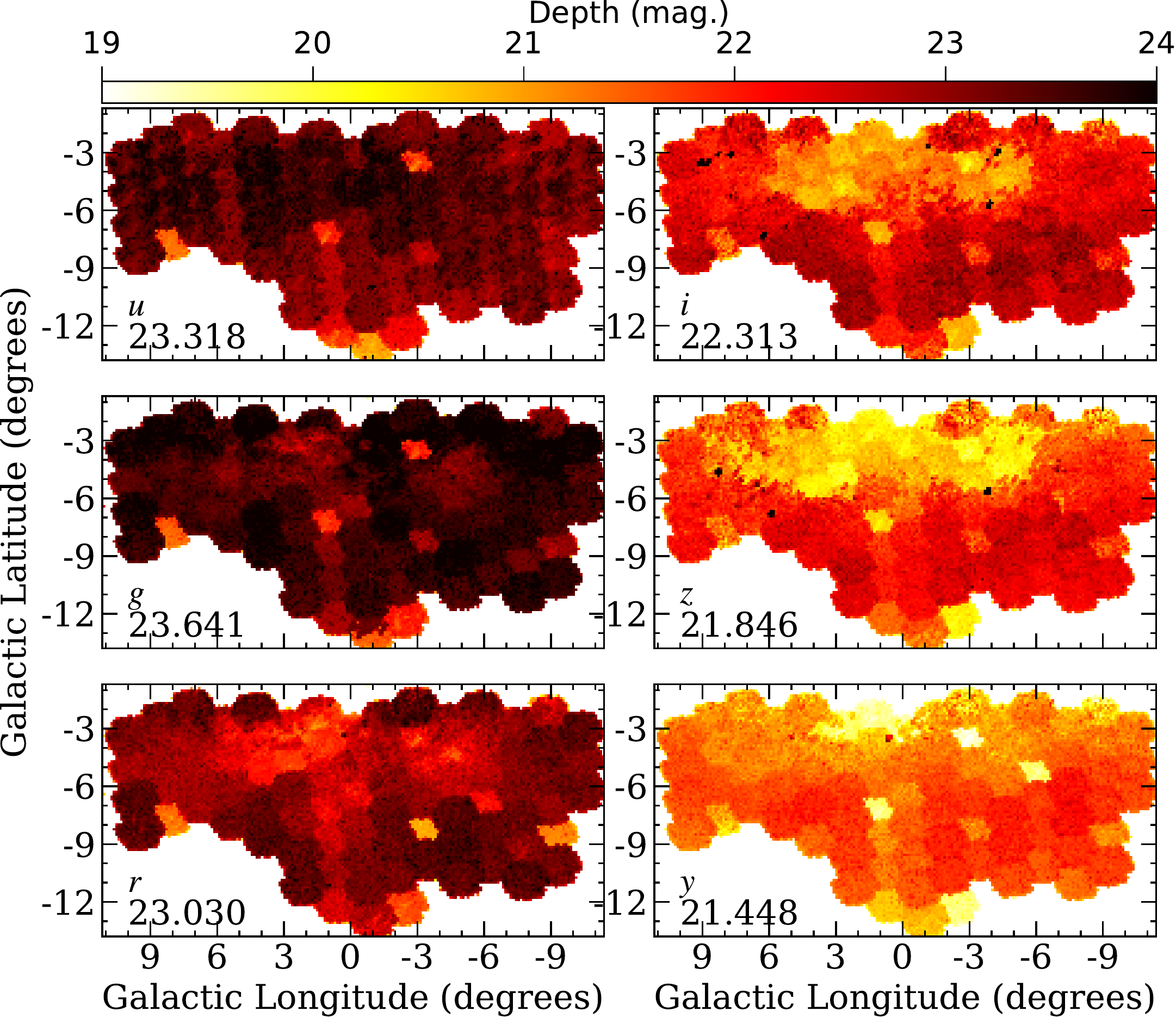}
\caption{Field depths are shown as a function of both filter and sky position.
The color gradient in each bin represents the $ugrizy$ magnitude of the
faintest star with a S/N ratio of at least 5.  Median image depths are provided
in each panel.}
\label{fig:depth_plot}
\end{figure}

The sky conditions were generally photometric for the 2013 runs, but the seeing
and cloud cover were highly variable during the 2014 run\footnote{Time-lapse 
video from the radiometric all-sky infrared camera \citep[RASICAM;][]{Lewis10} 
are available on Youtube for each night.}.  The seeing, based on the measured
FWHM of all $r$-band images, ranged from $\sim$ 0.8-1.8$\arcsec$, with an 
average of 1.1$\arcsec$ ($\sigma$ $\sim$ 0.2$\arcsec$).  The poor sky 
conditions in 2014 primarily affected the short and ultra-short exposures, 
which were a priority in the last run, but also limited the observing depth in 
some fields.  An example of the depth limitations can be seen in the 
($l$,$b$) = ($-$3,$-$8) field of Fig.~\ref{fig:density_map}, which has a much 
lower source density than surrounding fields.  The field-of-view was also 
reduced to 60 CCDs, instead of the nominal 61, for the 2014 run, which further 
affected the observational efficiency.  Nevertheless, Fig.~\ref{fig:depth_plot}
shows that a majority of the fields reached approximately the same depth in 
each filter, for a single long exposure.  Median calibrated 5$\sigma$ image 
depths for the $ugrizy$\footnote{Note that the $y$-band magnitudes listed in 
Fig.~\ref{fig:depth_plot} were calculated after the DECam $Y$ magnitudes were 
converted onto the $y$$_{\rm ps1}$ system.} bands are approximately 23.5, 23.8,
23.5, 23.1, 22.5, and 21.8 magnitudes, respectively.  Additional information 
regarding observing strategies, calibration fields (SDSS equatorial Stripe 82
with right ascensions of 10, 12, and 14 hours), and a catalog of observation 
dates for each field is provided in Rich et al. (2020, submitted).

All raw data products were processed using the NOAO Community Pipeline
\citep{Valdes14}, which automatically performs calibrations for issues such as:
overscan and bias correction, cross-talk, flat-fielding, fringe removal, 
pupil correction, and World Coordinate System (WCS) mapping.  Both the raw and
pipeline reduced images are available for download on the NOAO Science 
Archive\footnote{The NOAO Science Archive can be accessed at: 
http://archive1.dm.noao.edu/.}.  We note that all of the data presented here
are based on analyses of the calibrated images, and we did not incorporate the 
weight maps, data quality masks, or other Community Pipeline data products.

\section{Data Analysis and BDBS Pipeline Description}

\subsection{Data Management}
The data volume for all Community Pipeline processed BDBS science and 
calibration fields totaled $\sim$ 15 terabytes (TB) of space when uncompressed.
The number of utilized DECam images, which were obtained under acceptable but 
not necessarily photometric conditions and for which a successful WCS was 
found, totaled $\sim$ 7600, and is equivalent to $>$ 450,000 unique CCD images.
Given the large data volume and millions of files required to organize and 
analyze the DECam images, we stored and processed all BDBS data on Indiana 
University's (IU) 5 petabyte (PB) Data Capacitor II high speed shared storage 
system, which is managed by the Pervasive Technology Institute (PTI).  The IU 
system provided a central interface for the core BDBS investigators, which were
spread across various institutions, to process and share data products.
 
\begin{figure*}
\includegraphics[width=\textwidth]{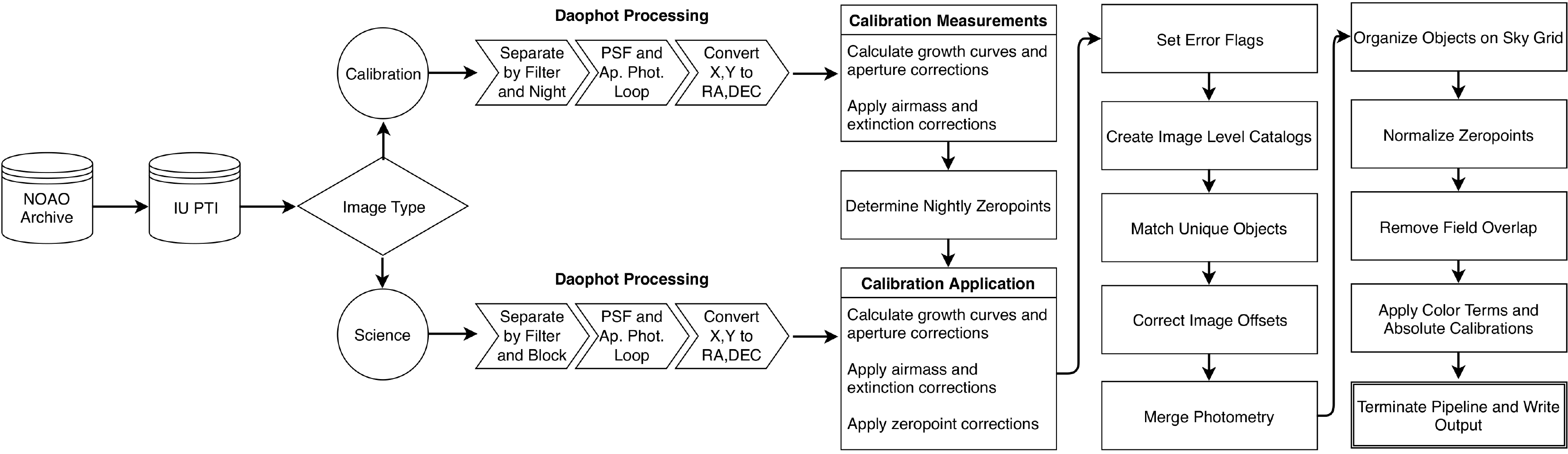}
\caption{A schematic diagram illustrating the data organization, data flow, and
general pipeline procedures.  Note that the final $u$-band calibration was
determined using DECam observations of the Sloan Digital Sky Survey 
\citep[SDSS;][]{Alam_2015} Stripe 82 field, but $g$, $r$, $i$, $z$, and $y$ 
were calibrated using direct overlap with the Pan-STARRS survey 
\citep{Chambers_2016}.}
\label{fig:bdbs_outline}
\end{figure*}

\begin{figure}
\includegraphics[width=\columnwidth]{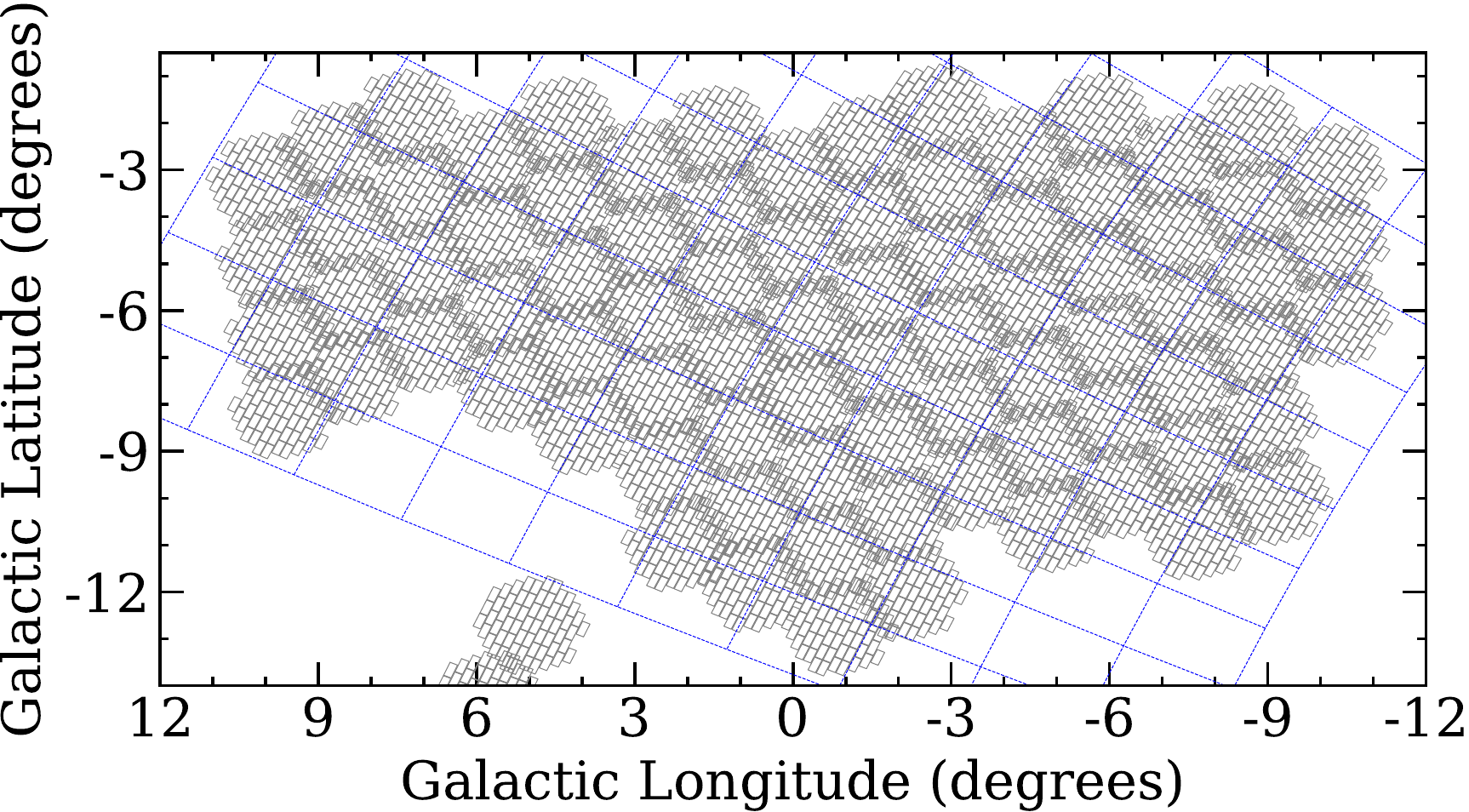}
\caption{The sky positions of individual CCD images are illustrated for the
 BDBS footprint with solid grey lines.  Each set of 60-61 CCDs
represents a unique pointing that included 2-15 additional small dithers for
each band (not shown).  The data were organized and photometered separately by
pointing and band, and were later organized onto the fixed sky grid shown by
the blue dashed lines.  The sky grid follows lines of constant right ascension
and declination.  Note that a few additional fields, such as the Sagittarius
dwarf spheroidal galaxy, were also observed for BDBS but are omitted because
they do not directly overlap with the contiguous portion of the survey.  We
anticipate including the Sagittarius dwarf photometry in the final data 
release.}
\label{fig:sky_grid}
\end{figure}

Fig.~\ref{fig:bdbs_outline} provides a basic outline of the data flow, file 
management, and processing scheme used for BDBS.  Briefly, the Community 
Pipeline calibrated images were transferred from the NOAO Science Archive to the
IU system via the parallel file transfer protocol tool provided by NOAO.  At
the highest level, images were partitioned into directories based on image type
(calibration or science).  Calibration frames were further separated based 
first on the night of observation and then by filter.  Similarly, science 
frames were organized first by field center and then by filter.  For reference,
Fig.~\ref{fig:sky_grid} shows the mean central position on the sky of each CCD 
within a given pointing ``block".

The multi-level data organization scheme has several practical and 
computational advantages.  For example, generating multiple directories based
on image metadata reduces the number of files within a subdirectory to a 
manageable level and speeds up operating system calls, such as file queries.
Additionally, an intuitive but descriptive file system structure provides a 
simple means for fast data archiving and compression, which is necessary for
both long and short term storage.  

Critical BDBS files that are not archived by NOAO, such as intermediate 
pipeline files, software products, and final catalogs, were placed into 
long-term storage on IU's 79 PB Scholarly Data Archive.  All data products were
temporarily stored on IU's Data Capacitor system for $\sim$ 1 year and accessed
via the Karst high-throughput computing cluster.  The final catalog release
paper (Johnson et al., in prep.) will describe additional long-term storage
and access solutions.

\subsection{Photometry Pipeline}
In order to take advantage of the distributed processing power of the IU 
supercomputing system, the mosaic DECam images mentioned in \S 3.1 were 
separated into individual CCD frames before entering the BDBS photometry
pipeline.  As illustrated in Fig.~\ref{fig:bdbs_outline}, all of the photometric
measurements were obtained using the point spread function (PSF) fitting 
DAOPHOT/ALLSTAR crowded field photometry software suite \citep{Stetson87}, as 
distributed by the Starlink Project\footnote{The Starlink software is currently
supported by the East Asian Observatory and can be accessed at: 
http://starlink.eao.hawaii.edu/starlink.} \citep{Currie14}.  The 
DAOPHOT/ALLSTAR interface was automated using a wrapper written in Fortran 2008
that handled all of the input/output (I/O) commands.  DAOPHOT/ALLSTAR was used
to perform actions such as: finding stars in an image, identifying acceptable
bright PSF stars ($>$100 per image for science fields; $>$10 per image for
Stripe 82 calibration fields), fitting a quadratically varying Gaussian PSF for
each 2K$\times$4K CCD, fitting and subtracting PSF models of stellar objects, 
and performing aperture photometry on images where all objects except the PSF
stars have been subtracted (for growth-curve analyses).  The general analysis 
procedure followed the recipe outlined in \citet{Stetson87}\footnote{See also 
http://www.star.bris.ac.uk/$\sim$mbt/daophot/mud9.ps.}, and involved three 
``fit and subtract" passes with DAOPHOT/ALLSTAR.  A final step in the pipeline 
converted the centroid X,Y pixel coordinates into right ascension and 
declination coordinates using the ``xy2sky" subroutine from WCSTOOLS 
\citep{Mink02}, which further utilized the WCS from the image headers.

\subsubsection{Photometry Pipeline Implementation}
The CCD images were processed through the BDBS pipeline on one of IU's high
performance shared storage systems.  Since the Fortran wrapper
was designed to run DAOPHOT/ALLSTAR independently on a single CCD image, a 
Python job generator script was constructed to scan all image folders and 
publish a job message to an Advanced Message Queuing Protocol (AMQP) queue 
system.  The AMQP job messages were shuffled so that the images were processed 
non-sequentially.  As mentioned previously, a major advantage of this method is
that different computing nodes were not simultaneously competing for access to 
the same folder on the shared storage system, which would otherwise be a time 
consuming I/O bottleneck.  Similarly, since the pipeline required certain 
static files in order to run (e.g., daophot.opt; allstar.opt; etc.), we avoided
name conflicts by processing all CCD frames in separate temporary 
subdirectories.

A set of 10 nodes, each with two 12-core Intel Xeon E5-2680 v3 processors, were
reserved on the Carbonate computing cluster at IU for the maximum allowable 
wall-time of 14 days.  On each node eight instances of a Python script were 
initiated that retrieved a job message from the aforementioned queue, executed 
the Fortran-based BDBS pipeline on a single CCD frame, logged usage statistics,
and then drew another job message from the queue.  Additional nodes were 
acquired when the initial allocation was consumed.  For the final production 
run, the BDBS pipeline required $\sim$ 3 weeks of wall-time and utilized a total
of $\sim$ 2.2 years of processor time on IU's Carbonate cluster.  Approximately
10$^{\rm 10}$ objects were identified when including all images and filters.

\subsubsection{Aperture and Nightly Zero Point Corrections}
At this point in the pipeline, the calibration and science fields contain an 
output file for each successfully run CCD frame that included metadata from the
analysis (e.g., FWHM; exposure time; filter and CCD ID; etc.) and for each 
star the measured RA/DEC coordinates, X/Y pixel positions, instrumental 
magnitude, standard magnitude error, local sky brightness estimate, and the 
``chi" and ``sharpness" parameters produced by DAOPHOT/ALLSTAR.  As mentioned 
previously, separate aperture photometry files are also included for all stars 
in the calibration frames but only the PSF stars in the science frames.  

For each unique combination of observation night and filter, the aperture
photometry files for both the science and calibration fields were copied into
temporary directories and processed with an implementation of the ``DAOGROW"
growth-curve algorithm described in \citet{Stetson90}.  This procedure produced
a mean aperture correction for each image that was applied to convert the
relative PSF magnitudes onto an absolute zero point scale.  Typical aperture
corrections were of order $-$0.3 magnitudes.  

\begin{figure}
\includegraphics[width=\columnwidth]{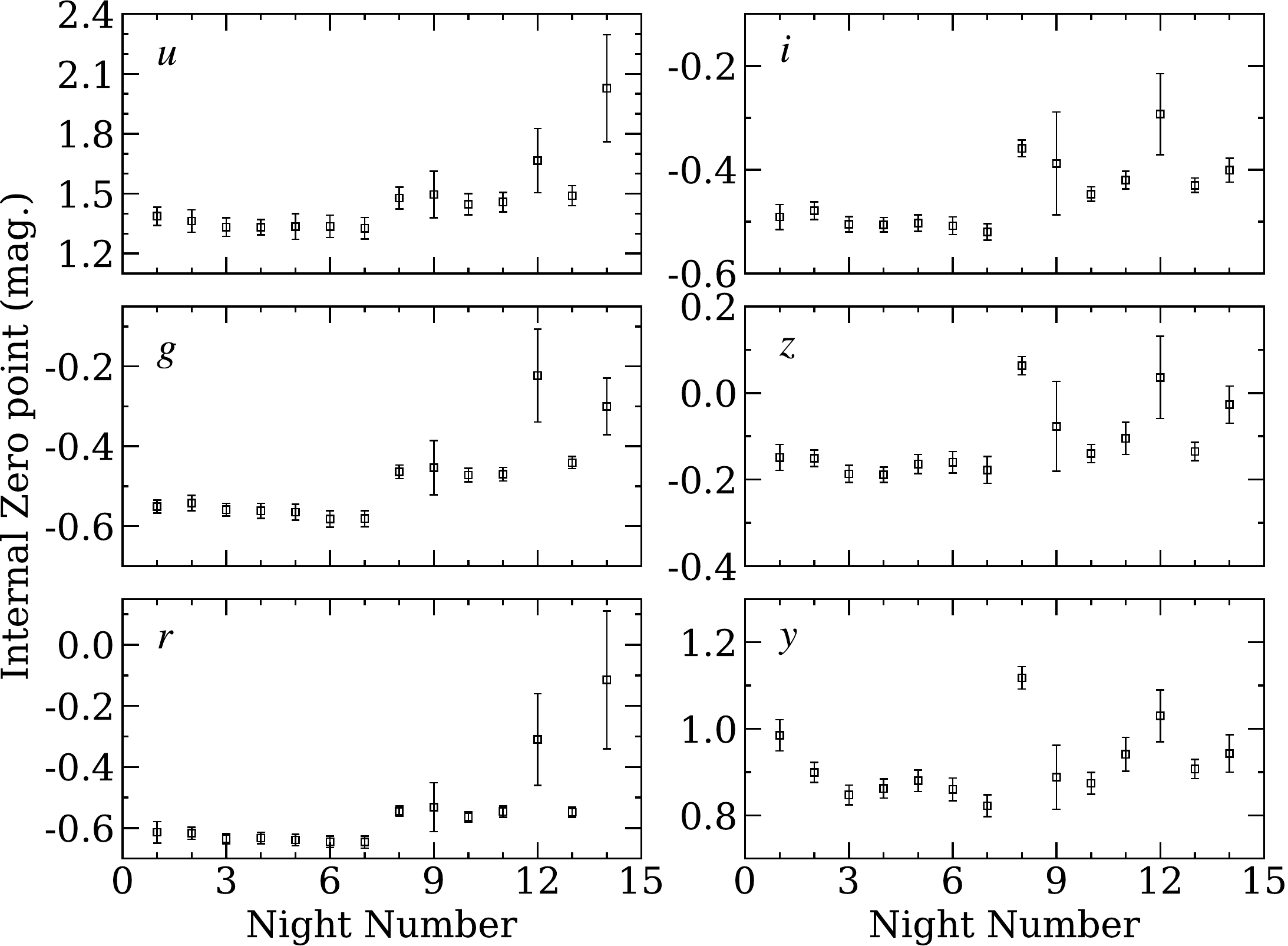}
\caption{The median internal zero point values, calculated using all CCDs and
exposures, are shown as open boxes for each filter of each observing night.
The error bars represent the median absolute deviations and are based on
sample sizes of $\sim$ 10,000 stars in the Stripe 82 fields.  Nights 1-5, 6-7,
and 8-14 span June 2013, July 2013, and June 2014, respectively.  Typical
median absolute deviation values are of order $\sim$ 0.02 mag., but during the
worst observing conditions the scatter reaches $\sim$ 0.1 mag.}
\label{fig:nightly_zp}
\end{figure}

Atmospheric extinction corrections were applied to all calibration and science 
instrumental magnitudes using the airmass values provided in the image headers 
and the extinction coefficients provided by NOAO\footnote{Estimates for 
photometric zero points, color corrections, and extinction corrections are 
provided at: http://www.ctio.noao.edu/noao/content/Mean-Photometric-Standard-Star-Module-PSM-Solutions-mean-zeropoints-color-terms-extinctions.}.  For the 
$ugriz$ filters, the Stripe 82 instrumental magnitudes were matched to the 
calibrated AB photometry from SDSS \citep{Alam_2015} while the $Y$-band data 
were matched to the UKIRT Infrared Deep Sky Survey Large Area Survey 
\citep[UKIDSS-LAS;][]{Lawrence07}.  Note that the UKIDSS-LAS $Y$-band data
were converted from $Y$$_{\rm VEGA}$ to $Y$$_{\rm AB}$ using the relation,
\begin{equation}
Y_{\rm AB} = Y_{\rm VEGA} + 0.634,
\end{equation}
from \citet{Hewett06}.  

Mean zero point values were iteratively calculated for each night and 
filter using the Stripe 82 calibration frames, and were applied to the 
appropriate aperture corrected science frame magnitudes.  
Fig.~\ref{fig:nightly_zp} illustrates the median internal zero 
point values for each observing night and filter, and clearly 
indicates that the 2013 observation conditions were superior to those of 2014. 
The application of these internal zero point corrections provided a blunt tool 
to correct for the different observing conditions on a night-to-night 
basis\footnote{One amplifier of the S7 CCD is unstable and has
poor linearity (see http://www.ctio.noao.edu/noao/node/2630), but as noted in
\citet{Schlafly_2018} the CCD nominally performs well.  Therefore, we did not
discard photometry from this CCD and treated the problematic amplifier the same
as all the others.  Erroneous photometry from this amplifier should be removed 
during the catalog merging process, but may persist in cases where only one
observation of a star on that amplifier was available.}.

\subsection{Catalog Generation and Final Calibration}
\subsubsection{Initial Catalogs}
Before merging the individual CCD results into image level catalogs (i.e., all
60-61 CCDs from a single DECam exposure), the pipeline automatically determined
simple quality control metrics based on a star's photometric measurement error,
local sky estimate, and chi value relative to other objects within a 0.05
mag. wide bin on the same chip in the same exposure.  For bins where the number
of objects was $<$ 100 (e.g., bright stars), the bin sizes were increased
until 100 targets were included.  Quality flags were set to indicate the number
of standard deviations away from the mean that each detection's error, sky,
and chi values resided.  Additional flags were added to the CCD level catalogs
to further indicate whether an exposure was long ($>$ 30 seconds) or short
($\leq$ 30 seconds) and whether it was obtained during a photometric or 
non-photometric night, as assessed from Fig.~\ref{fig:nightly_zp}.

For each exposure of each filter within a given pointing block (e.g., see 
Fig.~\ref{fig:sky_grid}), the CCD level catalogs were collated into image level
catalogs (i.e., all 60-61 CCD files of a single image were merged) in order to:
identify unique detections, measure and correct residual image-to-image zero 
point offsets, and generate a database of photometry measurements for each
unique source.  Before processing, the image level catalogs were indexed by
sorting via right ascension.  The coordinate sorting was accomplished by 
modifying a standard Fortran 2008 heapsort algorithm to operate on 
multi-dimensional arrays of varying data types, rather than just a 1D 
floating point array.

The adopted coordinate indexing scheme permitted the use of a bisecting ordered
search algorithm, which can quickly ($O$ $\sim$ log$_{\rm 2}$N) find 
upper/lower index boundaries in a 1D array when given minimum and maximum 
values.  We utilized this code to first produce a list of unique detections
within each pointing block and for each filter.  To generate the unique source 
list, a boolean array was constructed to flag whether an object had been 
previously matched to any other detection in the master list, for a given 
filter.  If a target of interest had not yet been discovered then the pipeline 
would scan the master catalog and perform a coordinate search with a radius of 
1$\arcsec$.  The bounding box for the coordinate search was analytically 
calculated with:
\begin{equation}
\delta_{\rm min.} = \delta_{\rm o} - r, 
\end{equation}
\begin{equation}
\delta_{\rm max.} = \delta_{\rm o} + r, 
\end{equation}
\begin{equation}
\alpha_{\rm min.} = \alpha_{\rm o} - \arcsin(sin(r)/cos(\delta_{\rm min.})), 
\end{equation}
\begin{equation}
\alpha_{\rm max.} = \alpha_{\rm o} + \arcsin(sin(r)/cos(\delta_{\rm max.})),
\end{equation}
where $r$ is the search distance, $\alpha$$_{\rm o}$ and $\delta$$_{\rm o}$ are 
the central right ascension and declination, and $\alpha$$_{\rm min.}$, 
$\alpha$$_{\rm max.}$, $\delta$$_{\rm min.}$, and $\delta$$_{\rm max.}$ are
the minimum and maximum right ascensions and declinations of the bounding 
boxes.  All objects inside the 1$\arcsec$ search radius were flagged as matches
to avoid double counting and unnecessary future searches, but only the closest
match was registered.  The catalog index values for all matches of unique 
objects were saved in a 2D matrix for rapid retrieval.  Coordinate statistics, 
such as median values, dispersions, and covariances, were calculated for each 
unique object.

\subsubsection{Intermediate Catalogs}
Separate intermediate catalogs were produced for each filter, and included the
internal photometric measurements and associated metadata for every unique
detection in a given band.  Since the observations are spread over two years 
and variable sky conditions, these intermediate catalogs were used to calculate
residual median zero point offsets between image pairs using well-measured 
bright (14-18 mag.) stars.  For each filter and sky block combination, a 
reference image was chosen to serve as the zero point anchor.  Priority for the
reference images was given first to long and then short exposures taken on 
photometric nights, primarily in 2013 (see Fig.~\ref{fig:nightly_zp}), long and
then short exposures taken on partially photometric nights, and then long or 
short exposures taken on non-photometric nights.  Each intermediate catalog 
maintained a flag to indicate the type of reference image used.

\begin{figure}
\includegraphics[width=\columnwidth]{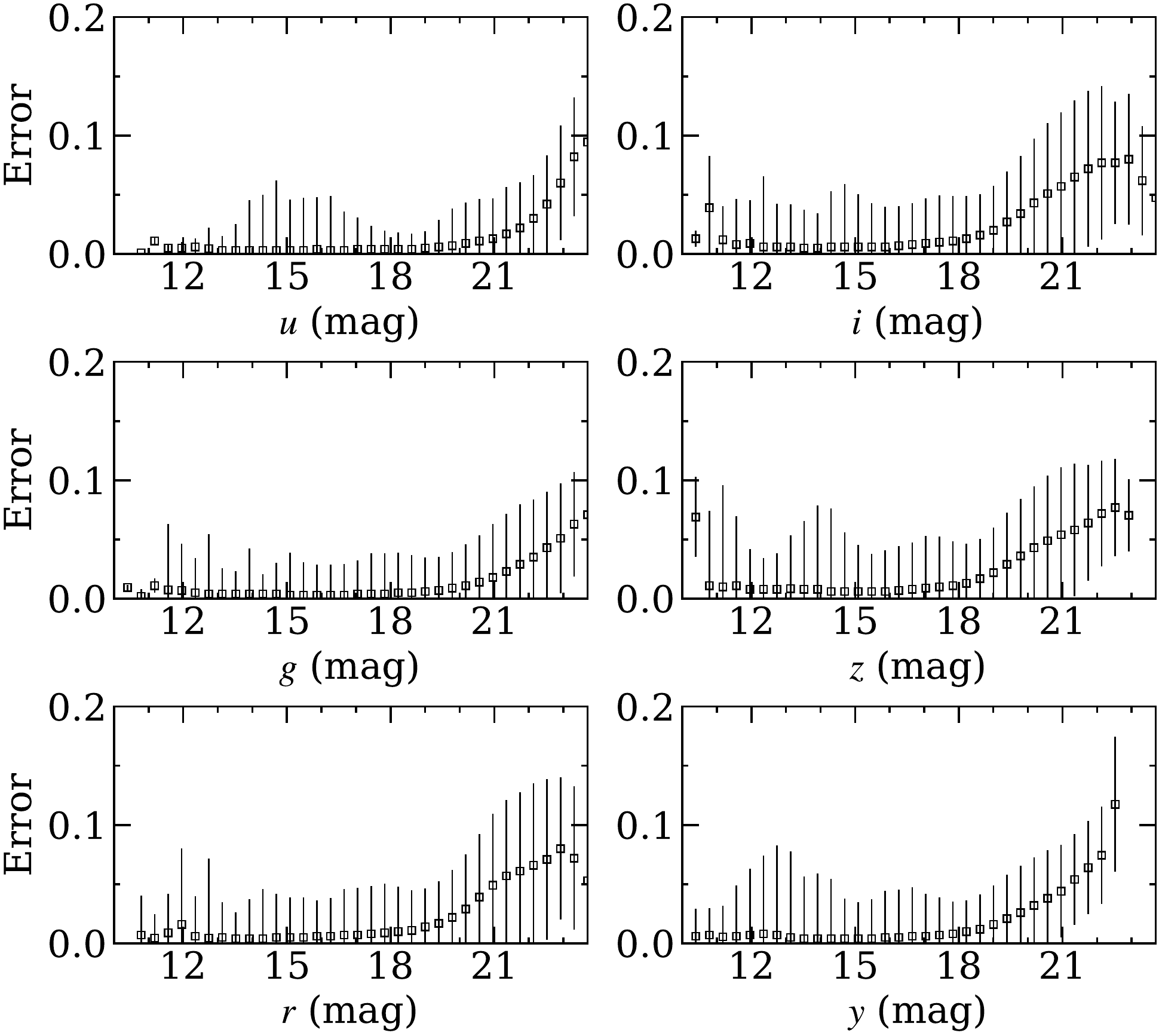}
\caption{Binned median photometric errors are shown for the $ugrizy$-bands
analyzed for BDBS.  The data are from a roughly 2 square degree field centered
near ($l$,$b$) = (0$\degr$,$-$6$\degr$) and represents a typical BDBS
pointing.  The error bars trace the median absolute deviations within each bin.
Note that the panels only include targets for which at least two clear
detections were present in a given band.  However, the total number of
observations can range from 2 to more than 15 for a given object.}
\label{fig:error_plot}
\end{figure}

The procedure described above ensured that for a given band all photometric
measurements shared a common, but not necessarily accurate, zero point.  
Therefore, all photometric measurements for each unique object in each filter
could be combined to reduce the observed scatter.  For objects with five or
more measurements in a given filter, a simple sigma clipping algorithm was
utilized to remove extreme outliers before combining\footnote{Note that our 
adopted measurement combination method is biased against variable sources.  
Objects with magnitudes that vary significantly over the course of hours to 
days, such as RR Lyrae, are underrepresented in our final catalogs.}.  However,
objects with four or fewer measurements included all values in the weighted 
average.  A weight array was determined for each photometric value using the 
inverse of the measurement variance.  A weighted measurement uncertainty was 
also determined along with a naive unweighted error value representing the 
uncertainty values from each image added in quadrature.  A summary of the 
weighted error distribution as a function of magnitude for all filters, but 
only including objects for which two or more detections were present, is 
provided in Fig.~\ref{fig:error_plot}.  The numerical error flags described in 
\S 3.3.1 were also combined using the same weight array to produce a simple 
measurement quality flag.  The number of combined measurements for each star 
was recorded and ranged from one to more than 15 for $>$ 400 million 
detections.

\subsubsection{Final Catalogs}
The intermediate cataloging phase identified unique objects within a dithered
pointing block, corrected the instrumental magnitudes for atmospheric 
extinction, applied a rough but uniform zero point correction to the AB system
\citep{Oke83}, and calculated a weighted average magnitude for each object.
However, Fig.~\ref{fig:sky_grid} shows that significant overlap exists between 
pointing blocks, which means that some stars are present in more than one 
intermediate catalog.  Furthermore, most intermediate catalog zero points were
anchored to a long exposure data set taken on a photometric night, but some 
were only anchored to long/short exposures taken on partially or 
non-photometric nights.  

Since not all objects are detected in every filter, we selected the $r$-band 
as a compromise astrometric filter (i.e., the final coordinates are based on 
the WCS from $r$-band images only and an object is only in the final catalog if
it has at least one clear $r$-band measurement)\footnote{Since the $r$-band
color correction also requires a $g$-band measurement, technically a star will
only appear in the final catalog if it has photometry from both $g$ and $r$.}.
For each pointing block, the unique object lists for each pass-band were 
filtered against the $r$-band source list using the same sort/search algorithms
described in \S 3.3.1.

\begin{figure}
\includegraphics[width=\columnwidth]{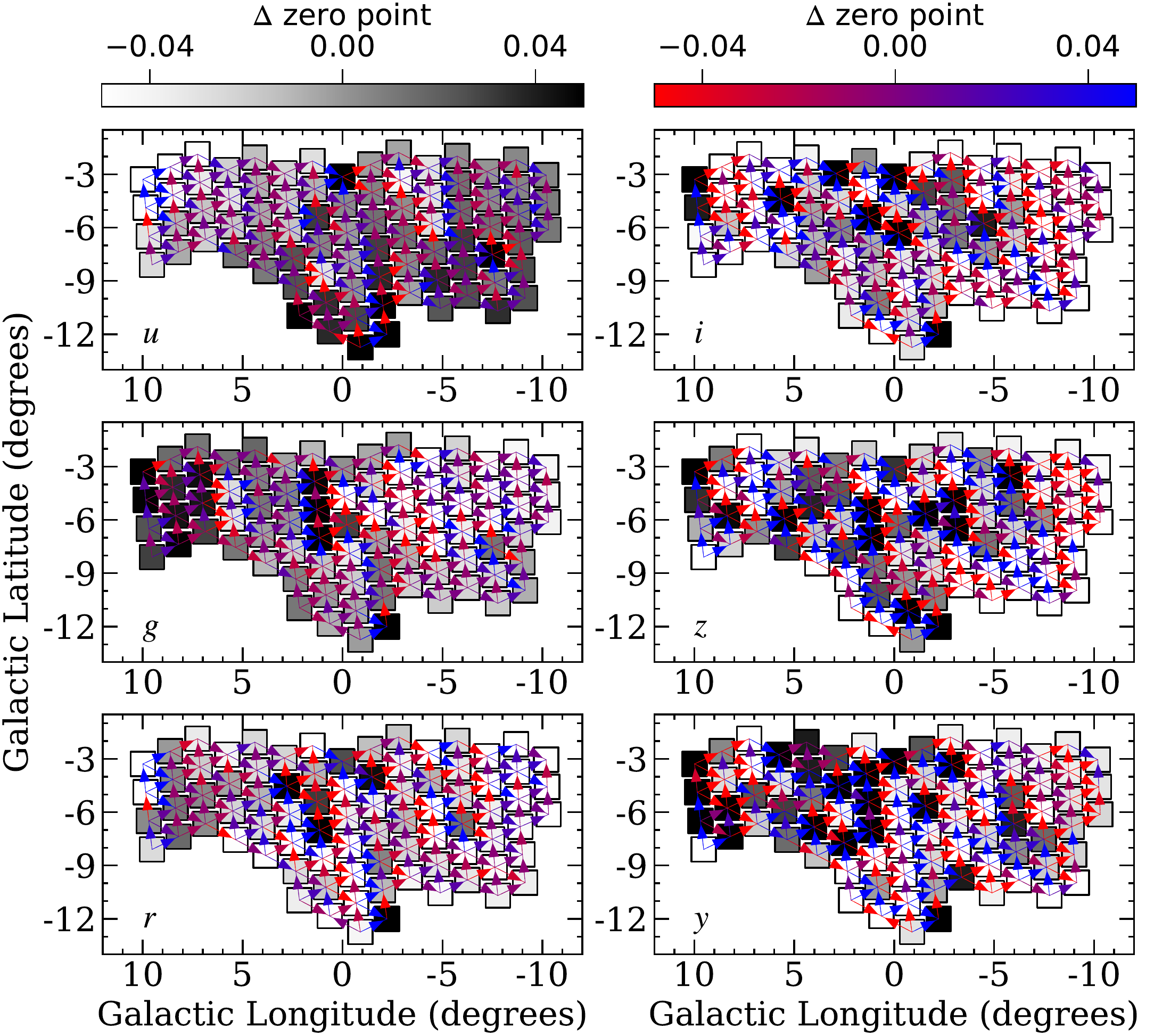}
\caption{Each box represents a unique pointing ``block" (e.g., see
Fig.~\ref{fig:sky_grid}) with the filled grey scale colors representing the
subtracted zero point corrections from the singular value decomposition.  The
field-to-field zero point offsets range from $-$0.05 mag. (white) to $+$0.05
mag. (black).  Since only a small number of fields in each filter had 
absolute corrections $>$ 0.05 mag., the color gradient was saturated at 
$\pm$0.05 mag.  The red ($-$0.05 mag.) to blue ($+$0.05 mag.) lines indicate 
the sign and magnitude of the offsets between individual fields.  The arrows 
illustrate the direction in which the zero point offsets were calculated.}
\label{fig:offset_grid}
\end{figure}

Before merging the photometry for detections that appear in two or more fields,
we used these objects to calculate the residual field-to-field zero point 
differences between adjacent fields.  The results of this process are 
illustrated by the colored arrows shown in Fig.~\ref{fig:offset_grid}, and
largely reflect differences in observing conditions between the reference 
frames described in \S 3.3.2.  Median field-to-field offsets and absolute 
deviations for the $ugrizY$ filters were $-$0.002 (0.019), $-$0.003 (0.019), 
$-$0.002 (0.026), $-$0.005 (0.035), $-$0.005 (0.064), and $-$0.003 (0.078) 
mag., respectively, and were based on an average of 2000 ($u$) to 80000 ($Y$) 
bright stars per field.

Fig.~\ref{fig:offset_grid} shows that all fields contain at least two adjacent
neighbors with object overlap and therefore a simple field-to-field offset 
correction is not practical.  Instead, we selected a pointing block where all 
observations in all filters were obtained on the same photometric night in 2013
to serve as a reference field.  A global solution for the field-to-field 
offsets in each filter that minimizes the zero point differences between all 
fields was then obtained via singular value decomposition.  The resulting zero 
point shifts applied to each field are illustrated by the filled boxes in 
Fig.~\ref{fig:offset_grid}.  Median absolute deviations for the applied
offsets in the $ugrizY$ filters were 0.020, 0.022, 0.018, 0.040, 0.036, and
0.046 mag., respectively.

With the entire data set now placed on a consistent zero point scale that is
ultimately tied to an exposure set obtained on a photometric night, objects
within each pointing block were allocated into the appropriate absolute 
sky grid bins illustrated in Fig.~\ref{fig:sky_grid}.  A final sort and search
pass was performed within each sky grid to match repeat observations (i.e.,
objects that appeared in two or more pointing blocks) and merge their 
photometry using the weighted averaging scheme described in \S 3.3.2.  This 
procedure produced a set of 77 catalogs where all photometry for a unique 
object has been merged and repeat observations have been excised.

\subsubsection{Absolute Calibration}
To complete the catalog collation process, color transformations and absolute 
zero point corrections were applied to the entire data set.  Although BDBS aims
to calibrate onto the AB system, no single reference data set is available that
runs from the near-UV to near-IR.  For example, SDSS only includes $ugriz$
whereas Pan-STARRS only has $grizy$.  Since a significant fraction of the BDBS
footprint, along with most of the utilized filters, overlays with Pan-STARRS,
we calibrated the BDBS data onto the Pan-STARRS $grizy$ system and reserved
the Stripe 82 SDSS calibration only for the $u$-band.

\begin{figure}
\includegraphics[width=\columnwidth]{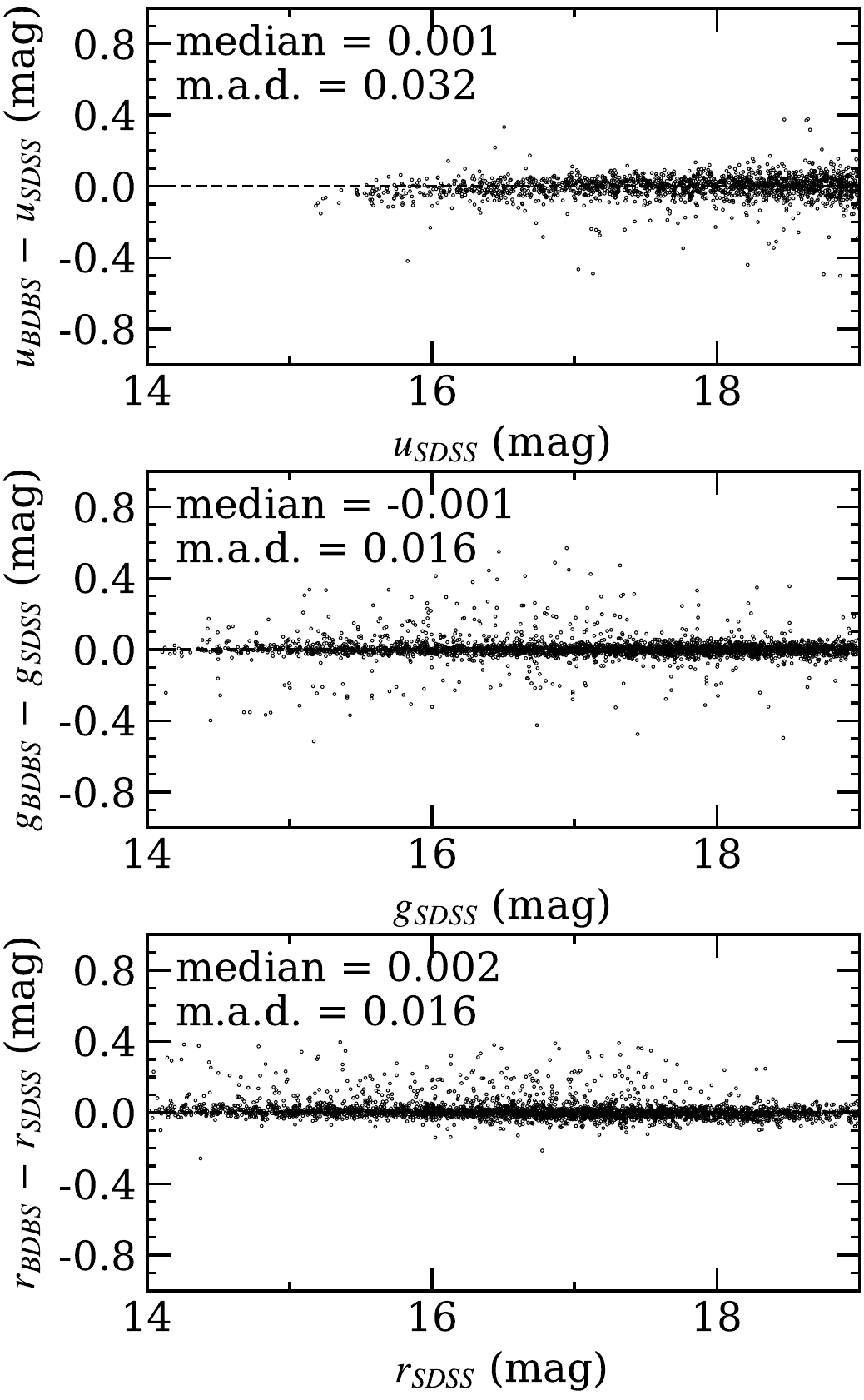}
\caption{The top panel shows the difference between the $u$$_{\rm BDBS}$ and
$u$$_{\rm SDSS}$ magnitudes as a function of $u$$_{\rm SDSS}$ for stars in
the Stripe 82 calibration field.  The Stripe 82 BDBS data were processed using
the same code and settings as the science fields (see
Fig.~\ref{fig:bdbs_outline}).  Since the conversion from the internal magnitude
system to the SDSS system included a color term based on $u-g$, which also
depends on $g-r$, similar comparisons are shown in the middle and bottom
panel for the $g$ and $r$ bands.  Note that the calibration equations used to
transform the internal $g$ and $r$ magnitudes onto the SDSS system were only
used for the purposes of the $u$-band calibration.  The final $g$ and $r$
magnitudes were calibrated onto the Pan-STARRS system instead.  The median
magnitude differences and median absolute deviations (m.a.d.) are provided in
all three panels.}
\label{fig:u_calib}
\end{figure}

The $u$-band calibration was relatively straight-forward and utilized the 
Stripe 82 observations from the same photometric night as the internal zero 
point frame described in \S 3.3.3.  The $u$-band calibration from the internal
BDBS system to SDSS utilized the $u-g$ color term, which also required the
$g-r$ term, and was calculated using:
\begin{equation}
g_{\rm BDBS} = g_{\rm int.} - [-0.109(g-r)_{\rm SDSS} + 0.056],
\end{equation}
\begin{equation}
r_{\rm BDBS} = r_{\rm int.} - [-0.079(g-r)_{\rm SDSS} + 0.042],
\end{equation}
\begin{equation}
u_{\rm BDBS} = u_{\rm int.} - [0.008(u-g)_{\rm SDSS} - 0.010],
\end{equation}
where u$_{\rm BDBS}$ represents the calibrated $u$-band magnitude in the 
final BDBS catalog, u$_{\rm int.}$ is the internal uncalibrated magnitude,
and ($u-g$)$_{\rm SDSS}$ is the iteratively determined $u-g$ color corrected
to the SDSS system.  Note that equation 8 is relevant for stars with 0.75 $\la$
($u-g$)$_{\rm SDSS}$ $\la$ 2.75 mag.  Residuals between the final calibrated
u$_{\rm BDBS}$ magnitudes and those in the SDSS catalog for the Stripe 82 
calibration field are shown in Fig.~\ref{fig:u_calib}.  The median absolute 
deviation for stars with magnitudes between 15-19 is 0.032 mag.  Similar 
comparisons are also provided in Fig.~\ref{fig:u_calib} for the $g$ and $r$
filters.  Since the $u$-band data are calibrated using the SDSS data from
\citet{Alam_2015}, an equivalent shift of 0.02 mag. may need to be added to
the BDBS $u$-band data to place it on the formal AB system\footnote{For further
information we refer the interested reader to http://classic.sdss.org/dr7/algorithms/fluxcal.html.}.

\begin{figure}
\includegraphics[width=\columnwidth]{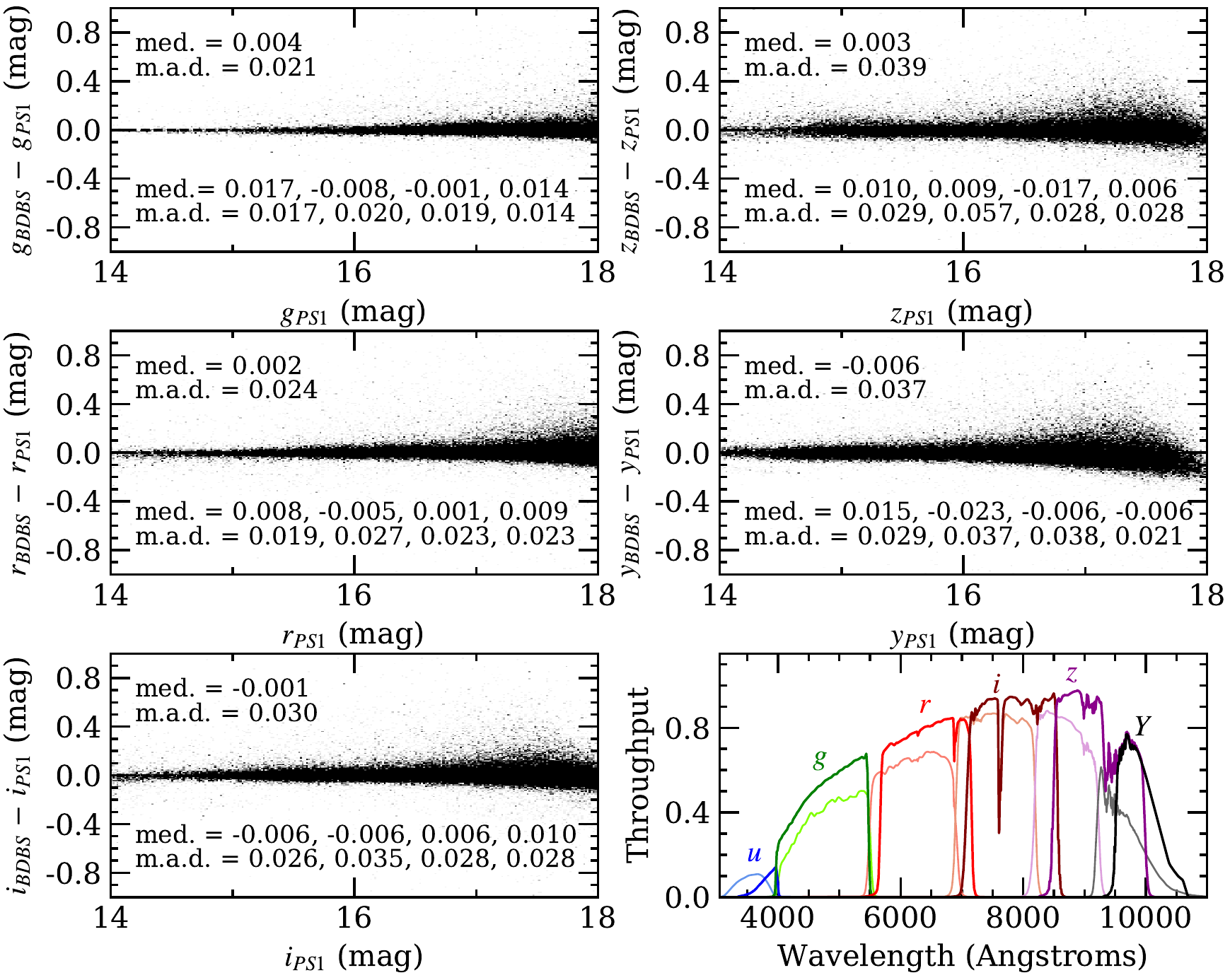}
\caption{Similar to Fig.~\ref{fig:u_calib}, photometric differences between
BDBS and Pan-STARRS are plotted as a function of Pan-STARRS magnitudes for the
$grizy$ bands.  The panels include four fields located at ($l$,$b$) $\approx$
($+$3.5,$-$6.5), ($+$6.0,$-$5.5), ($+$7.0,$-$4.5), and ($+$9.0,$-$3.5), which
were used for determining the calibration equations between the internal BDBS
system and Pan-STARRS.  The top median and median absolute deviation values
include all four fields and the bottom sets indicate the statistics for
individual fields.  The bottom right panel shows throughput curves for the
DECam $u$ (blue), $g$ (green), $r$ (red), $i$ (maroon), $z$ (dark magenta), and
$Y$ (black) filters in comparison with the $u$$_{\rm SDSS}$ and $grizy$$_{\rm
PS1}$ filters, which are shown as lighter colors.}
\label{fig:grizy_calib}
\end{figure}

As mentioned previously, the $grizy$ filters were calibrated directly using
overlap between Pan-STARRS and BDBS.  In particular, we selected the zero point
reference field from \S 3.3.3 and three adjacent fields that were also observed
on completely photometric nights as the calibration sets.  Since only the 
$g$ and $r$ filters are guaranteed to exist for all stars in the final catalog,
we utilized $g-r$ as the reference color for the transformations between the
internal BDBS $grizY$ and Pan-STARRS $grizy$ systems.  By comparing bright
($<$ 18 mag.) overlapping BDBS and Pan-STARRS detections in four reference 
fields, we determined the following transformation equations:
\begin{equation}
g_{\rm BDBS} = g_{\rm int.} - [-0.055(g-r)_{\rm PS1} - 0.039],
\end{equation}
\begin{equation}
r_{\rm BDBS} = r_{\rm int.} - [-0.095(g-r)_{\rm PS1} - 0.042],
\end{equation}
\begin{equation}
i_{\rm BDBS} = i_{\rm int.} - [-0.079(g-r)_{\rm PS1} - 0.106],
\end{equation}
\begin{equation}
z_{\rm BDBS} = z_{\rm int.} - [-0.087(g-r)_{\rm PS1} - 0.124],
\end{equation}
\begin{equation}
y_{\rm BDBS} = Y_{\rm int.} - [-0.050(g-r)_{\rm PS1} - 0.082],
\end{equation}
where similar to equation 8 the left hand side of equations 9-13 represent the
final BDBS magnitudes calibrated onto the Pan-STARRS system and the right hand
side of equations 9-13 include the internal uncalibrated magnitudes and the
calibrated ($g-r$)$_{\rm PS1}$ colors.  The calibrations above are relevant for
stars with ($g-r$)$_{\rm PS1}$ colors between about $-$0.1 and 2.0 mag.  As
noted in \citet{Schlafly_2018}, additional offsets of order 0.020, 0.033, 
0.024, 0.028, and 0.011 may need to be added to the calibrated BDBS magnitudes
to place them on the absolute AB scale.

A summary of the residuals between the calibrated BDBS magnitudes and the 
Pan-STARRS catalog for overlapping objects in all four calibration fields is
provided in Fig.~\ref{fig:grizy_calib}.  Note that a comparison between the
filter transmission curves of DECam and SDSS/Pan-STARRS is provided in
Fig.~\ref{fig:grizy_calib} as well.  The individual median offsets for each of 
the four fields, along with the median absolute deviations, is also provided.  
The overall median absolute deviations for the $grizy$ filters are 0.021, 
0.024, 0.030, 0.039, and 0.037, respectively, and the field-to-field absolute 
calibration variance is generally $\la$ 0.01-0.02 mag.  We adopt the overall
median absolute deviations from Fig.~\ref{fig:grizy_calib} as an estimate for 
the calibration uncertainty for the $grizy$ filters.  The final, calibrated 
BDBS catalog contains 243,959,076 unique objects and will be released in a 
future publication (Johnson et al., in prep.).

\subsection{Reddening and Extinction Corrections}
Fig.~\ref{fig:density_map} clearly shows that many BDBS lines-of-sight 
suffer from significant small and large-scale differential extinction due to
foreground dust.  Therefore, we corrected for reddening using the high sampling
(1$\arcmin$$\times$1$\arcmin$) extinction map from \citet{Simion17}, which
covers most of the BDBS footprint ($\vert$$l$$\vert$ $<$ 10$\degr$; 
$-$10$\degr$ $<$ $b$ $<$ $+$5$\degr$).  The \citet{Simion17} map was built 
using red clump giants from the VVV survey following the methods described in
\citet{Gonzalez11_ext,Gonzalez12}, which provided the first VVV extinction maps
but at a coarser (2$\arcmin$$\times$2$\arcmin$ to 6$\arcmin$$\times$6$\arcmin$)
spatial resolution.

Since the \citet{Simion17} map is derived from the VISTA photometric system,
several transformations were required to obtain $ugrizy$ extinction 
corrections.  The VISTA $JK_{\rm S}$ photometry was first converted to the 
2MASS system using Equations C4 and C6 from 
\citet{GonzalezFernandez18}\footnote{Additional information is provided by the
Cambridge Astronomy Survey Unit (CASU) at: http://casu.ast.cam.ac.uk/surveys-projects/vista/technical/photometric-properties.  Note that we utilized version 
1.3 of the VISTA data.}, and the reddening values were converted using the
relation:
\begin{equation}
E(J-K)_{\rm 2MASS} = 1.081E(J-K)_{\rm VISTA}.
\end{equation}
These transformations were then combined with Table 1 of \citet{Green18} to 
determine E(B$-$V) as:
\begin{equation}
E(B-V) = 2.045E(J-K)_{\rm 2MASS}.
\end{equation}
Finally, the extinction values for each filter were computed as:
\begin{equation}
A_{\rm u} = 4.239E(B-V),
\end{equation}
\begin{equation}
A_{\rm g} = 3.384E(B-V),
\end{equation}
\begin{equation}
A_{\rm r} = 2.483E(B-V),
\end{equation}
\begin{equation}
A_{\rm i} = 1.838E(B-V),
\end{equation}
\begin{equation}
A_{\rm z} = 1.414E(B-V),
\end{equation}
\begin{equation}
A_{\rm y} = 1.126E(B-V),
\end{equation}
using Table 6 in \citet{Schlafly11} for the $u$-band and Table 1 of 
\citet{Green18} for the $grizy$-bands.  

\begin{figure}
\includegraphics[width=\columnwidth]{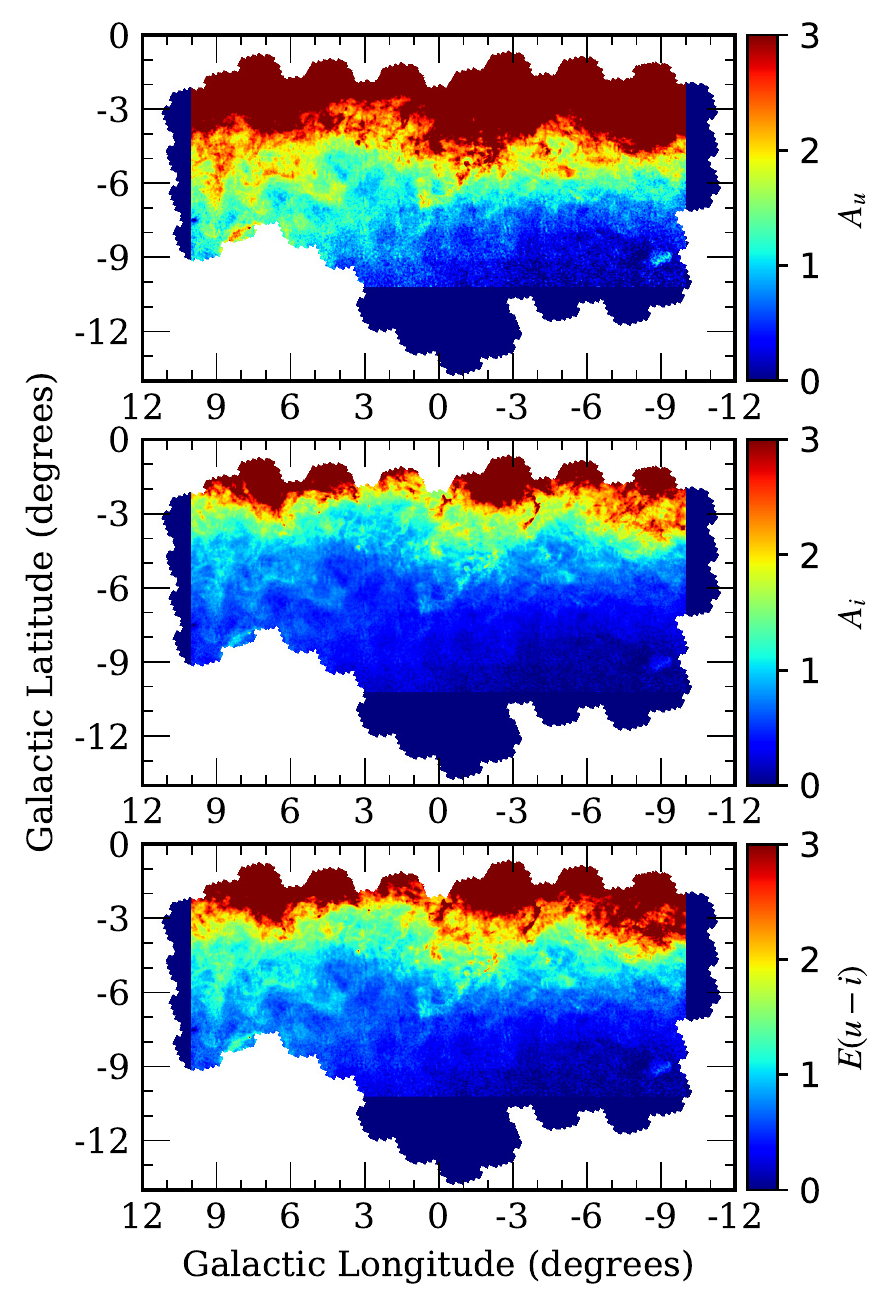}
\caption{The top, middle, and bottom panels illustrate sample A$_{\rm u}$,
A$_{\rm i}$, and E($u-i$) extinction and reddening maps used for BDBS.  
Equivalent maps exist for the $grzy$ filters as well.  The extinction and 
reddening values are based on the bulge maps presented in \citet{Simion17}.
Note that the extinction and reddening values are only valid for fields with
$\vert$$l$$\vert$ $<$ 10$\degr$ and $b$ $>$ $-$10$\degr$.}
\label{fig:ext_map}
\end{figure}

For the $grizy$-bands, the adopted \citet{Green18} extinction vector was
based on a combination of broad band stellar colors and APOGEE spectra 
\citep[see also][]{Schlafly16}.  Since a majority of the APOGEE reference stars
used to define the extinction vector reside within the disk and bulge, the 
adopted vector is appropriate to use for BDBS.  For the $u$-band, the 
\citet{Schlafly11} extinction vector was obtained by comparing the colors of 
synthetic spectra built using stellar parameters from the SEGUE Stellar 
Parameter Pipeline \citep{Lee08}.  By construction, the extinction vector for 
the $u$-band is more appropriate for higher Galactic latitudes.  However, in 
the absence of a more recent determination in high extinction regions we use 
the $u$-band extinction law for R$_{\rm V}$ = 3.1 in Table 6 of 
\citet{Schlafly11}.  We caution that a small error in the reddening law, or a 
small variation in the reddening law between different lines-of-sight, will
produce large errors in the $u$-band extinction correction.

Extinction maps for the $u$-band and $i$-band, along with a sample E($u-i$) 
reddening map, are provided in Fig.~\ref{fig:ext_map}.  Note that the 
extinction and reddening maps from \citet{Simion17} are only valid for BDBS 
stars with $\vert$$l$$\vert$ $<$ 10$\degr$ and $b$ $>$ $-$10$\degr$.  
Reddening corrections are not available at this time for fields with higher 
longitudes and/or latitudes.  Additionally, a recent analysis by 
\citet{Hajdu19} showed that spatially varying zero point offsets of order 
0.05-0.1 mag. in E($J-K_{s}$) are present in the VVV data.  However, the 
largest zero point deviations are limited to fields within $\sim$ 2$\degr$ of 
the plane, where BDBS observations are sparse, and \citet[][see their 
Section 2]{Simion17} already corrected the VVV photometry for many of the 
effects noted in \citet{Hajdu19}.

\section{Data Validation and Early Science Results}

\subsection{A Comparison with DECaPS and Pan-STARRS}

As noted in $\S$ 1, several recent surveys have employed large format imagers
to observe low latitude disk and bulge fields in optical
band passes.  However, few of these surveys specifically aim to understand the
formation history and structure of the Galactic bulge.  For example, Pan-STARRS
is geared toward obtaining uniform $grizy$ photometry for all regions of the
sky with $\delta$ $\ga$ $-$30$\degr$.  On the other hand, the DECaPS and VST 
Photometric H$\alpha$ Survey \citep[VPHAS+;][]{Drew14} programs use various
combinations of the $grizy$ (DECaPS) and $ugriH\alpha$ (VPHAS+) filters to
primarily understand the Galactic disk.  \citet{Saha19} also used the 
$ugriz$ filters on DECam to explore the Galactic bulge, but this work 
emphasizes RR Lyrae and focuses on a small number of fields that span $\sim$
1 per cent of the BDBS footprint.  In this sense, BDBS is unique because the 
project's core goal is understanding $>$ 200 contiguous square degrees of the 
Galactic bulge by utilizing uniform photometry spanning the $ugrizy$ filters.

\begin{figure}
\includegraphics[width=\columnwidth]{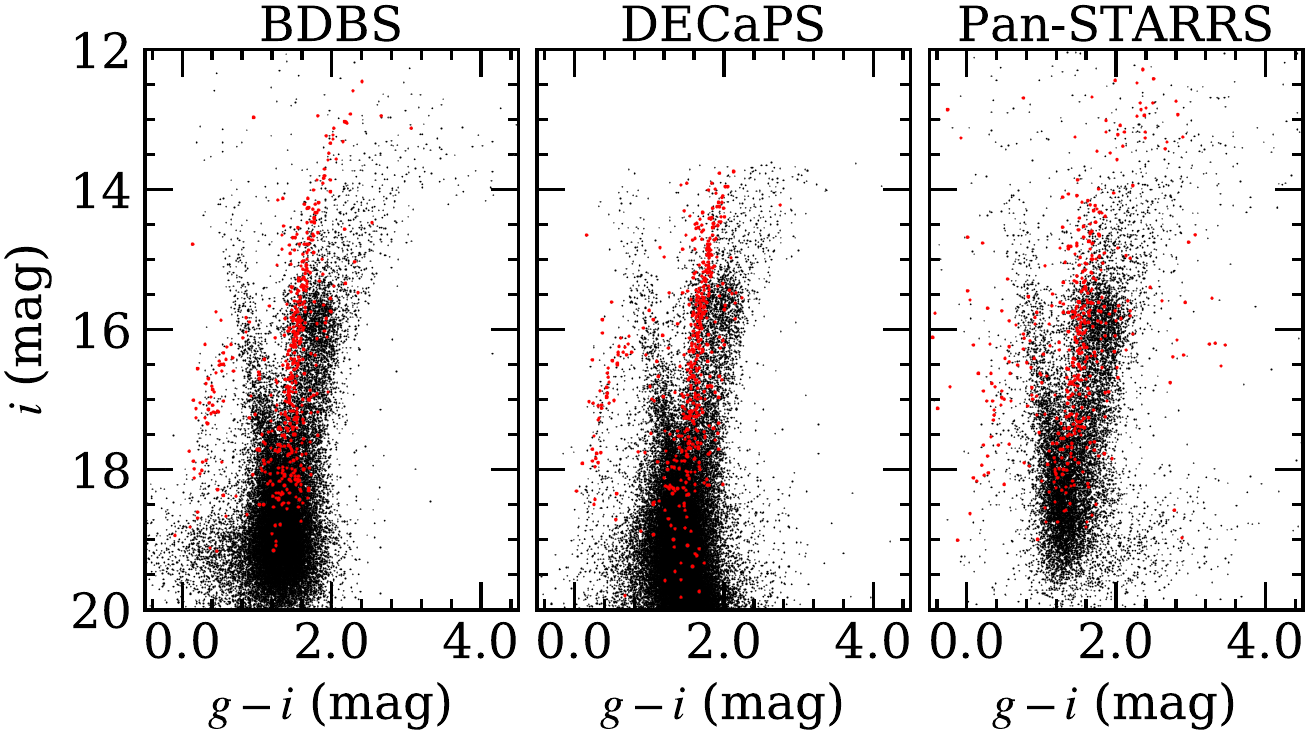}
\caption{The three panels show $i$ versus $g-i$ CMDs for 5$\arcmin$ regions
near the Baade's window globular cluster NGC 6522 using data from BDBS (left),
DECaPS (middle), and Pan-STARRS (right).  The small black points represent
field stars while those shown as filled red points have Gaia DR2 proper motions
consistent with NGC 6522 membership \citep{Gaia18_GCs}.  At least for this 
field, the DECaPS $i$-band data are systematically brighter by $\sim$ 0.2 mag. 
than BDBS and Pan-STARRS.}
\label{fig:survey_comp}
\end{figure}

However, before proceeding to analyze the BDBS CMDs we first seek to verify
data quality using the DECaPS and Pan-STARRS catalogs for comparison.  
Fig.~\ref{fig:survey_comp} compares $i$ versus $g-i$ CMDs of a region within
5$\arcmin$ of the globular cluster NGC 6522 for the BDBS, DECaPS, and 
Pan-STARRS catalogs.  This particular line-of-sight also resides within the 
well-studied ``Baade's window" region, which is a low extinction window near
($l$,$b$) $\sim$ ($+$1,$-$3.9).  Although Baade's window is a commonly analyzed
field for bulge studies, it did not receive any special attention during the
calibration procedures for BDBS, Pan-STARRS, or DECaPS.  Therefore, the data
presented in Fig.~\ref{fig:survey_comp} should be a fair comparison between
the various studies of an effectively random field, which is useful for 
examining calibration uniformity.

Fig.~\ref{fig:survey_comp} shows that all three surveys produce morphologically
similar CMDs.  For example, the blue foreground disk sequences are relatively
tight and well-separated from the bulge RGB population, but merge with the 
bulge CMD sequence near $i$ $\sim$ 18 mag.  Additionally, all three 
CMDs show a prominent bulge red clump near $i$ $\sim$ 16 mag. along with a 
broad RGB that results from a combination of metallicity, distance, and 
reddening variations.  Similarly, the NGC 6522 stars, identified by Gaia DR2
proper motions \citep{Gaia18_GCs}, clearly reside on the blue edge of the bulge
RGB sequence, which is consistent with the cluster's comparatively low 
metallicity \citep{Barbuy09,Barbuy14,Ness14_6522}.  The cluster's blue 
horizontal branch (HB) population is also evident at $g-i$ $<$ 1 mag. in all 
three CMDs, and a similar group of field blue HB stars can also be seen.

\begin{figure}
\includegraphics[width=\columnwidth]{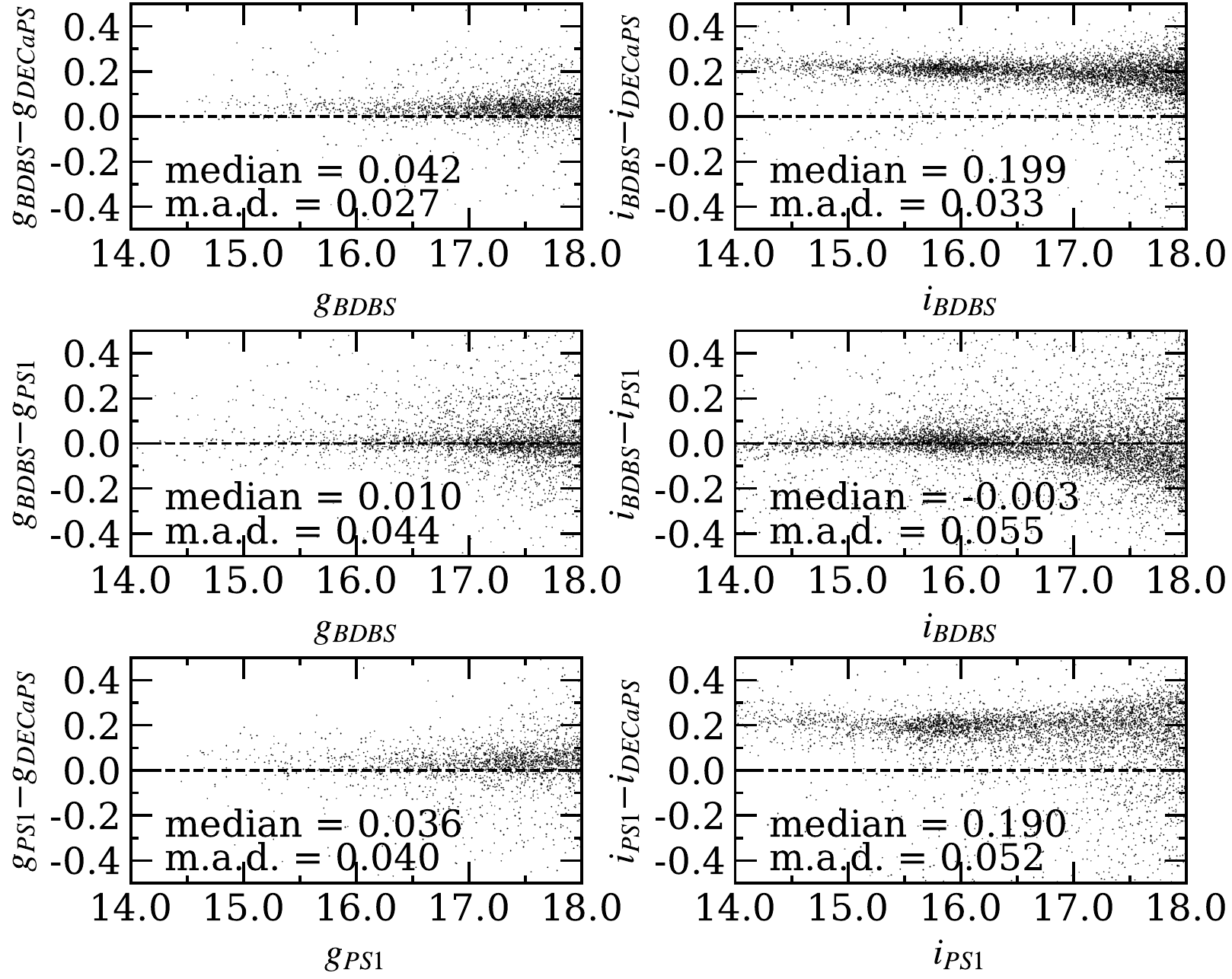}
\caption{Residual $g$ and $i$-band differences as a function of magnitude are
shown for bright stars observed in Fig.~\ref{fig:survey_comp}.  For this
particular field, the $g$ and $i$-band zero points are consistent between BDBS
and Pan-STARRS (PS1).  However, the DECaPS $g$ and $i$ data appear to be
systematically brighter by $\sim$ 0.04 and 0.19 mag., respectively.  The BDBS
and DECaPS data also exhibit some minor structure in the residual plots, which
is likely driven by imperfect color corrections.  The median offsets and
median absolute deviations are provided for each panel.}
\label{fig:survey_comp2}
\end{figure}

Despite these similarities, Fig.~\ref{fig:survey_comp} highlights clear 
differences between the surveys.  The Pan-STARRS data do not reach as deep as
BDBS and DECaPS, and the photometric scatter among NGC 6522 stars is noticeably
worse.  The DECaPS data reach $\sim$ 0.5 mag. deeper in $i$ compared to BDBS, 
but since DECaPS did not obtain short exposures the survey has a fainter bright 
limit compared to BDBS.  DECaPS and BDBS exhibit similar photometric precision
for the NGC 6522 and field stars that range from $i$ $\sim$ 14-20 mag., but
the DECaPS red clump appears brighter than those of BDBS and Pan-STARRS.  The 
DECaPS zero point offset is surprising given that both BDBS and DECaPS are 
calibrated off of Pan-STARRS.  However, \citet{Schlafly_2018} note that the 
DECaPS photometry is systematically brighter relative to Pan-STARRS in more
heavily reddened fields.

\begin{figure}
\includegraphics[width=\columnwidth]{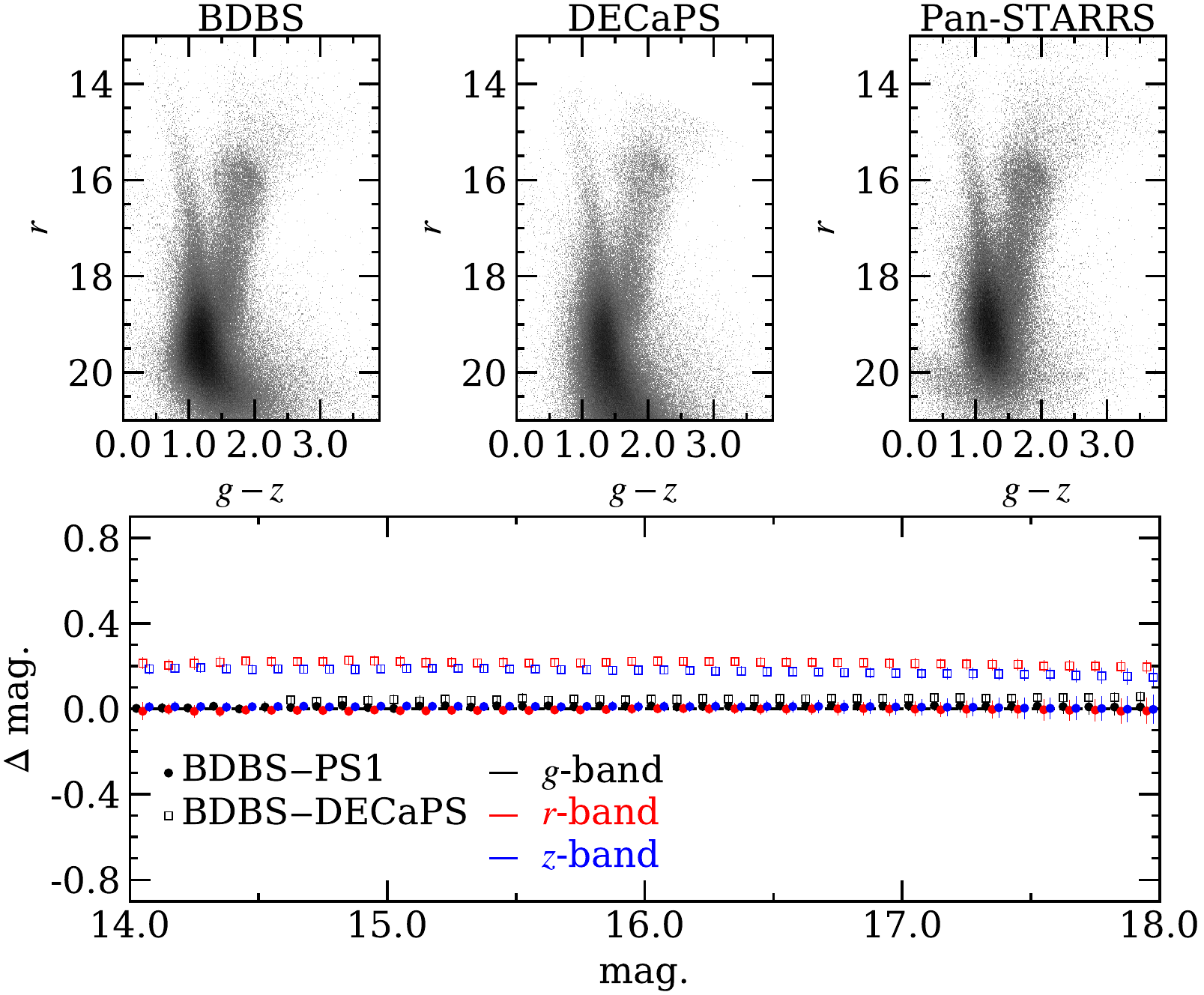}
\caption{Similar to Fig.~\ref{fig:survey_comp}, $r$ versus $g-z$ CMDs are 
compared between BDBS, DECaPS, and Pan-STARRS for a 20$\arcmin$ field near
($l$,$b$) = ($+$2,$-$6).  Residuals between the three surveys for each filter
are shown in the bottom panel with the black symbols illustrating the $g$-band
residuals, the red symbols showing the $r$-band residuals, and the blue symbols
showing the $z$-band residuals.  Since the residuals between BDBS and 
Pan-STARRS are close to zero, we have omitted the residual points between 
DECaPS and Pan-STARRS for figure clarity.  Note that the symbols represent the 
median offsets between the surveys in 0.1 mag. bins, and that the vertical 
error bars indicate the median absolute deviations.  A comparison between
BDBS/Pan-STARRS and DECaPS data shows a systematic offsets of $\sim$ 0.05-0.20 
magnitudes, depending on the filter.}
\label{fig:survey_comp5}
\end{figure}

Fig.~\ref{fig:survey_comp2} shows that for stars with magnitudes ranging from 
14-18 mag., the median offsets between DECaPS and Pan-STARRS are 0.036 mag.
and 0.190 mag. for the $g$ and $i$-bands, respectively.  Similarly, the 
$g$ and $i$-band offsets between BDBS and DECaPS are 0.042 and 0.199 mag.,
respectively.  By extension the median offsets between BDBS and Pan-STARRS 
are relatively small at 0.010 mag., and $-$0.003 mag. for $g$ and $i$, 
respectively.  Although the median absolute deviations between all three
studies do not exceed about 0.05 mag., Fig.~\ref{fig:survey_comp2} shows that
both BDBS and DECaPS exhibit a residual systematic offset of $\la$ 0.1 mag.
for stars with $i$ $\ga$ 17.5 mag.  This residual offset is likely due to higher
order terms in the color corrections, which can also be seen in 
Fig.~\ref{fig:grizy_calib}.  

Fig.~\ref{fig:survey_comp5} shows a similar comparison between BDBS, DECaPS, 
and Pan-STARRS in a less extincted and crowded field near ($l$,$b$) = 
($+$2,$-$6) but for the $grz$ filters, and indicates that the BDBS/Pan-STARRS 
photometry are again in good agreement.  The DECaPS data are systematically 
brighter than BDBS and Pan-STARRS by $\sim$ 0.2 mag. for the $r$ and $z$-bands 
but are within $\sim$ 0.05 mag. for the $g$-band.  Although we have only 
compared two random fields here for data exploration purposes, the evidence 
suggests that the zero point correction procedure illustrated in 
Fig.~\ref{fig:offset_grid} did an adequate job ``flattening" the data across 
the BDBS footprint.  

\subsection{Bulge CMDs and the Double Red Clump}

\begin{figure*}
\includegraphics[width=0.7\textwidth]{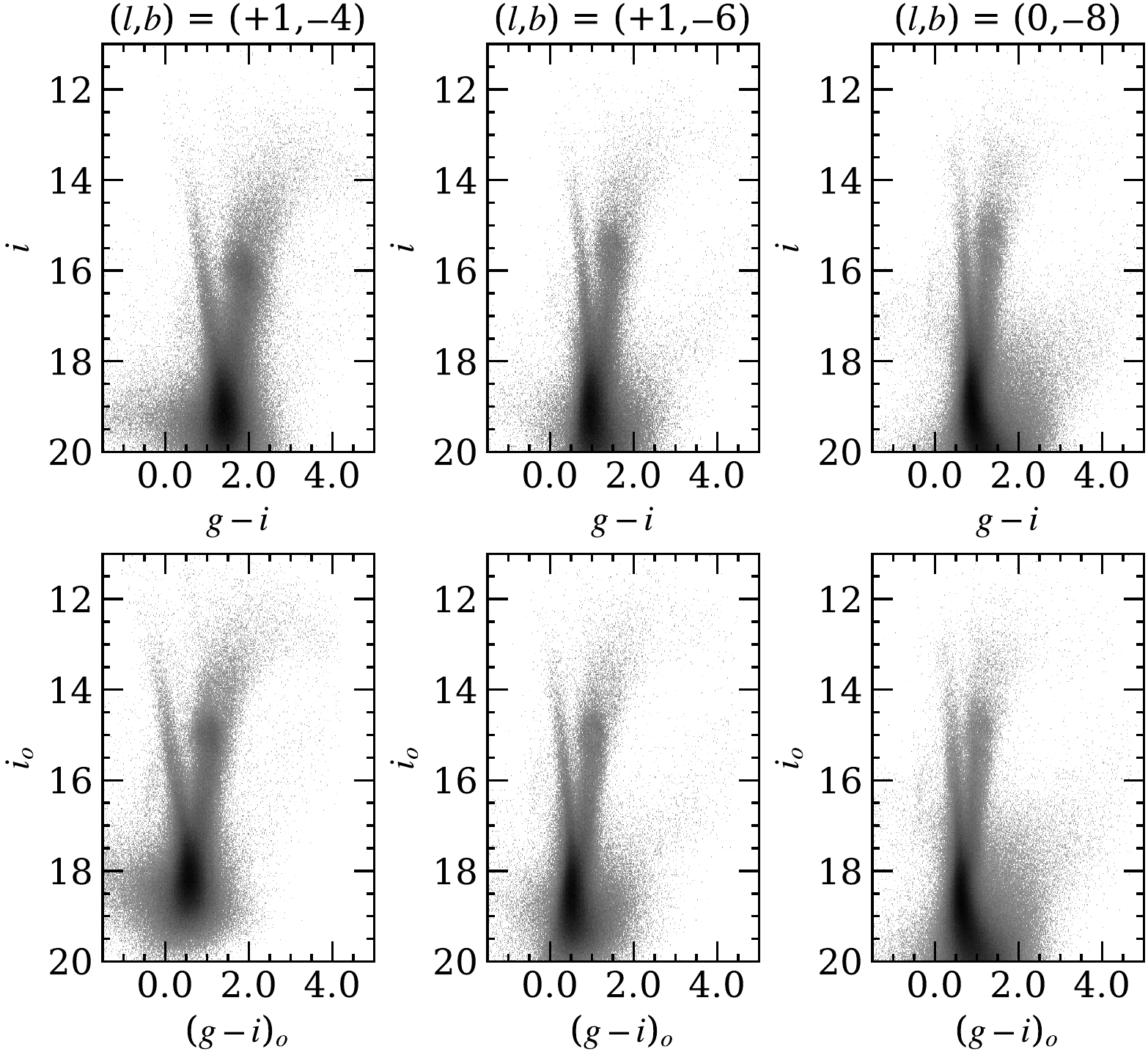}
\caption{The top panels show binned $i$ versus $g-i$ CMDs for three minor
axis fields at varying Galactic latitudes.  The bottom panels illustrate the
same fields but with reddening corrections from \citet{Simion17} applied (see
text for details).  Note that the $b$ = $-$4$\degr$ and $-$6$\degr$ fields span
a radial distance of 20$\arcmin$ from the noted field centers while the
$b$ = $-$8$\degr$ field spans a 40$\arcmin$ radius.  All three fields
show evidence of blue HB populations, and double red clumps are visible
in the $b$ = $-$6$\degr$ and $-$8$\degr$ fields.  The two left panels are 
comparable to Figure 10 in \citet{Saha19}.}
\label{fig:sample_cmds}
\end{figure*}

Galactic bulge CMDs can be difficult to interpret due to a variety of effects,
including large metallicity spreads, significant differential reddening, and
complex line-of-sight geometries related to the inner disk and bar.  Increased
reddening and crowding also conspire to decrease photometric depth, especially
in optical CMDs, for bulge fields close to the plane.  These effects are 
clearly illustrated in the sample BDBS CMDs of Fig.~\ref{fig:sample_cmds} where
we note that the Baade's window ($b$ = $-$4$\degr$) data reach about 1 mag.
below the main-sequence turn-off while the minor axis fields at $b$ $\la$ 
$-$6$\degr$ reach another 1-2 magnitudes fainter.  Fig.~\ref{fig:sample_cmds} 
also shows that the apparent magnitude of the red clump is faintest for Baade's 
window and steadily becomes brighter as the foreground extinction decreases at 
higher Galactic latitudes.

After applying the reddening corrections illustrated in Fig.~\ref{fig:ext_map},
as is shown in the bottom panels of Fig.~\ref{fig:sample_cmds}, we note that
all of the CMDs improve.  For example, the red clump apparent magnitude 
variations disappear, and other important regions, such as the main-sequence
turn-off and blue edge of the RGB, now reside at consistent ($g-i$)$_{\rm o}$ 
colors.  Additionally, the lower RGBs form tighter sequences and the blue HB
populations exhibit reduced photometric scatter.  The $b$ = $-$4$\degr$ panel
of Fig.~\ref{fig:sample_cmds} also shows the ``blue loop" population noted
by \citet{Saha19} at ($g-i$)$_{\rm o}$ $\sim$ 0.75 mag. and 
$i$$_{\rm o}$ $\sim$ 12.5 mag.  However, we show in Rich et al. (2020, 
submitted) that these stars have distances of $\sim$ 3 kpc or less, and are 
therefore likely foreground red clump stars.  

Reddening corrected CMDs, such as those shown in Fig.~\ref{fig:sample_cmds},
also permit more detailed morphological examinations.  As a verification case, 
we can explore the red clump regions of the three minor axis fields.  Previous 
studies such as \citet{McWilliam10} and \citet{Nataf10} showed that several
bulge sight lines with $\vert$$b$$\vert$ $>$ 5$\degr$ exhibit two red clumps
separated by $\sim$0.5 mag. in the I and K$_{\rm S}$-bands.  As mentioned 
in $\S$ 1, these and other authors contend that the double red clumps trace out
an ``X"-shaped structure, and that the two clumps reside at the near ($\sim$ 
6.5 kpc) and far ($\sim$ 8.8 kpc) ends of the bar.  However, other explanations
relating to chemical composition variations, particularly involving He 
enhancements, have also been suggested \citep[e.g.,][]{Lee15,Joo17,Lee18,Lee19,
Lopez19}.

\begin{figure*}
\includegraphics[width=\textwidth]{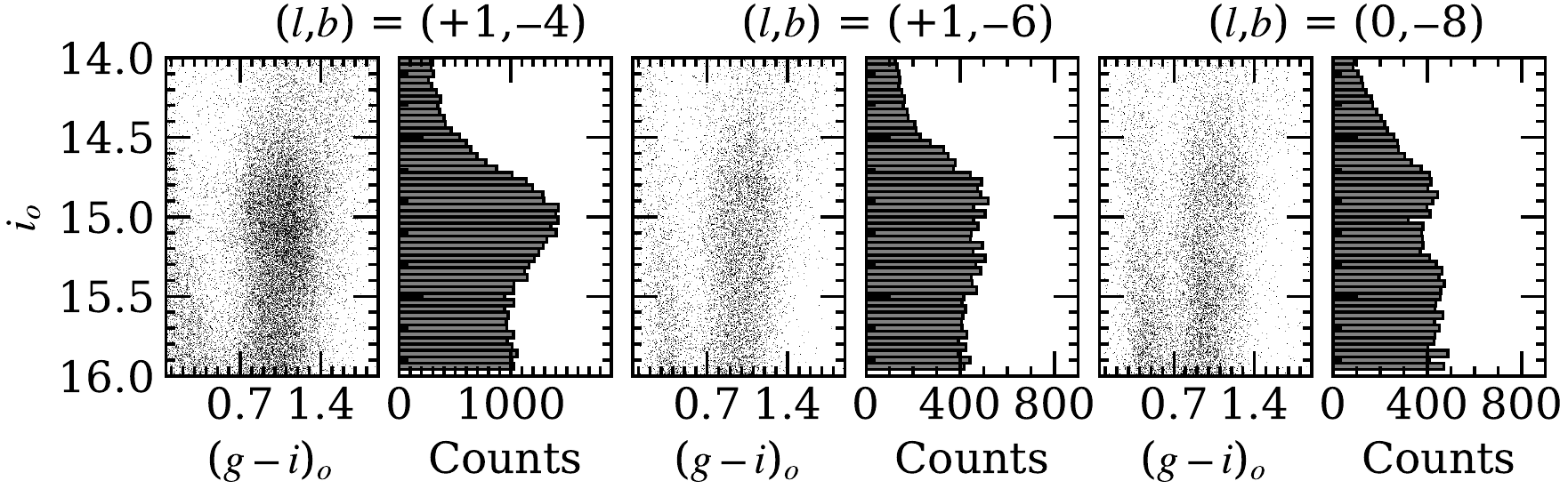}
\caption{Dereddened $i$ versus $g-i$ CMDs are shown for the same fields as
in Fig.~\ref{fig:sample_cmds} with an emphasis on the red clump region.
Histograms for each field illustrate the monolithic red clump population in
Baade's window that contrasts with the double red clumps, located at
$i$$_{\rm o}$ $\sim$ 14.9 mag. and 15.35 mag., found in higher Galactic
latitude fields.}
\label{fig:double_red_clump}
\end{figure*}

Fig.~\ref{fig:double_red_clump} highlights the red clump regions in dereddened
CMDs for the same minor axis fields shown in Fig.~\ref{fig:sample_cmds}.  As
expected, the minor axis field near $b$ = $-$4$\degr$ exhibits a strong
but unimodal red clump at $i$$_{\rm o}$ $\sim$ 15.0 mag.  However, the 
outer bulge fields at $b$ = $-$6$\degr$ and $-$8$\degr$ exhibit clear double
red clumps separated by $\sim$ 0.4-0.5 mag. in the $i$-band.  This separation
is similar to that observed by \citet{Nataf10} using OGLE observations.  For
the $b$ = $-$6$\degr$ field, the bright and faint red clumps peak near
$i$$_{\rm o}$ = 14.9 and 15.3 mag., respectively, but the two clumps are found
near $i$$_{\rm o}$ = 14.9 and 15.4 mag. in the $b$ = $-$8$\degr$ field.  The
increased brightness separation for higher latitude bulge fields is consistent
with the 2MASS K$_{\rm S}$-band measurements from \citet{McWilliam10}.

\begin{figure*}
\includegraphics[width=\textwidth]{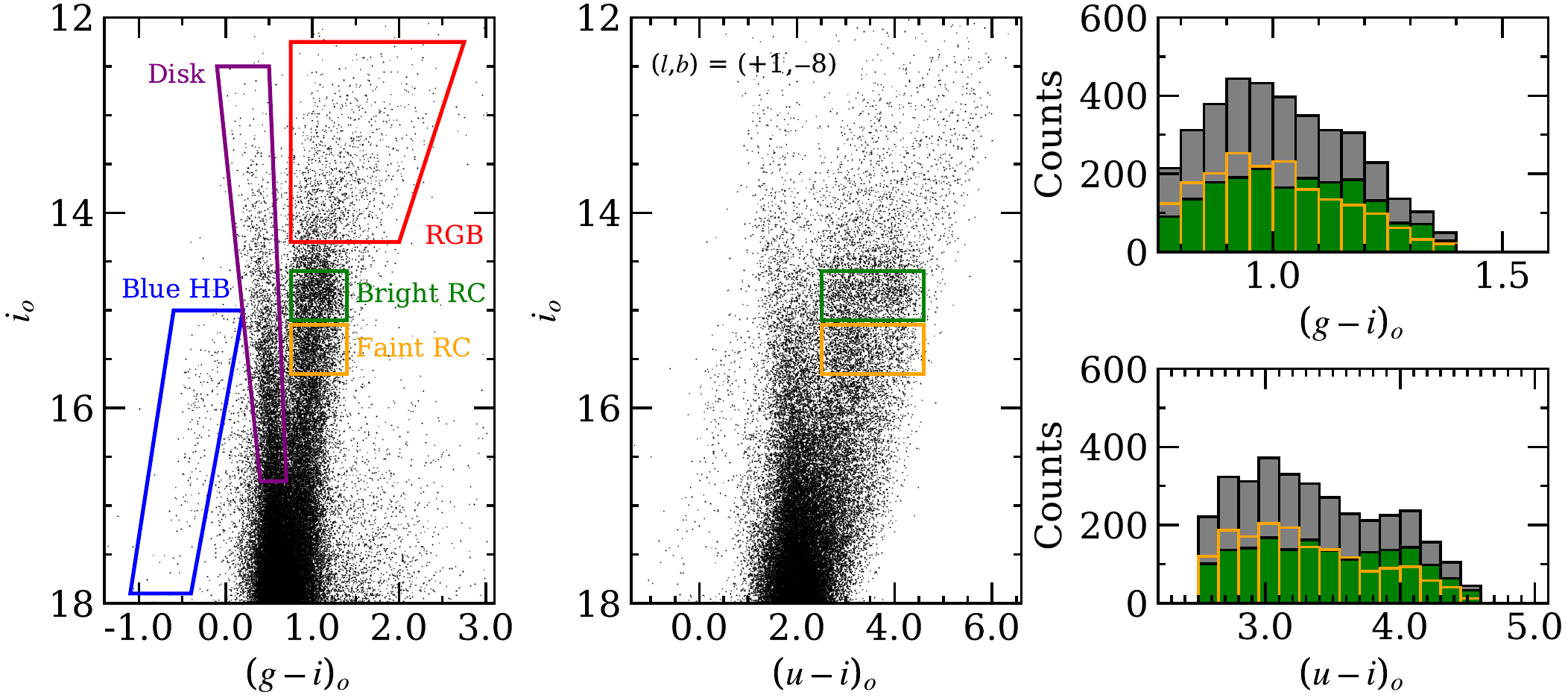}
\caption{\emph{Left:} a dereddened $i$ versus $g-i$ CMD for a minor axis
field at ($l$,$b$) = ($+$1,$-$8) with a radius of 20$\arcmin$.  Several CMD
features are highlighted, including the RGB, extended blue HB, foreground disk,
and bright/faint red clump populations.  \emph{Middle:} A similar CMD using
($u-i$)$_{\rm o}$ as the color base line.  The green and orange boxes
indicate the locations of the bright and faint red clumps, respectively.  Note
that the bright and faint red clumps each contain separate populations of
``blue" and ``red" populations.  \emph{Right:} the top and bottom panels show
histograms of the red clump populations as functions of ($g-i$)$_{\rm o}$
and ($u-i$)$_{\rm o}$, respectively.  The grey histograms include all red
clump stars while those shown in green and orange represent the bright and
faint red clumps, respectively.  Note the clear bimodal distribution present
when using ($u-i$)$_{\rm o}$.}
\label{fig:split_rc}
\end{figure*}

A more detailed examination of the $b$ = $-$8$\degr$ field is shown in 
Fig.~\ref{fig:split_rc}.  The foreground disk, blue HB, double red clumps,
and bulge RGB sequences are all well-populated.  The blue HB is particularly
extended and ranges from at least $i$$_{\rm o}$ = 15-18 mag.  Similar
features are also present in DECam CMDs involving the $rzy$ filters, as is
evident in Figures 9 and 11 from \citet{Saha19} and those shown in $\S$ 4.4 of
the present work.  Fig.~\ref{fig:split_rc} also shows that the red clump 
($g-i$)$_{\rm o}$ dispersion is relatively small, and that the bright and faint
red clumps effectively span the same color range.  Interestingly, we find some 
weak evidence of bimodality in the ($g-i$)$_{\rm o}$ red clump histogram,
particularly for the bright red clump, which could be indicative of a 
bimodal metallicity distribution.  However, Fig.~\ref{fig:split_rc} shows that
the $u$-band is a much stronger filter for separating stars by chemical
composition.

\subsection{The $u$-band and Metallicity Distribution Functions}

\subsubsection{Red Clump Color Split}

The most striking feature of Fig.~\ref{fig:split_rc} is that adding the 
$u$-band dramatically increases the color dispersion of bulge stars on the 
RGB, which is particularly evident in the red clump region.  When plotted using
a $g-i$ baseline, the double red clump spans $\sim$ 0.6 magnitudes in color.
However, with $u-i$ the two red clumps span $>$ 2 magnitudes in color.  
Furthermore, the addition of the $u$-band splits each component of the double
red clump into two populations that are separated by about 1 mag. in $u-i$
(making four in total).  The separate blue and red populations are clearly seen
as distinct clumps in the $i$$_{\rm o}$ versus ($u-i$)$_{\rm o}$ panel of 
Fig.~\ref{fig:split_rc}, but the ratios of the blue/red populations vary 
between the bright and faint red clumps.  

At least for the ($l$,$b$) = ($+$1,$-$8) sight line shown in 
Fig.~\ref{fig:split_rc}, the bright red clump has a blue fraction of 0.622 
$\pm$ 0.039 and a red fraction of 0.378 $\pm$ 0.046, when separated at 
($u-i$)$_{\rm o}$ = 3.65 mag.  Similarly, the faint red clump has blue 
and red fractions of 0.747 $\pm$ 0.038 and 0.253 $\pm$ 0.055, respectively.  
A similar discrepancy between the bright and faint red clumps has been noted by 
previous authors as well \citep{Ness12,Uttenthaler12,Rojas17}, which found the
bright red clump to contain a larger fraction of metal-rich stars compared
to the faint red clump distributions.

\subsubsection{Linking the $u$-band with Metallicity}

The color split on the RGB, and particularly in the red clump region, is a 
common feature seen in the BDBS $i$$_{\rm o}$ versus ($u-i$)$_{\rm o}$ CMDs.
Furthermore, the observed CMDs (i.e., not corrected for reddening) in outer 
bulge fields where the extinction is much lower (e.g., see 
Fig.~\ref{fig:ext_map}) are morphologically similar to the dereddened 
versions, and suggest that the $u-i$ red clump color dispersions and discrete 
populations, such as those seen in Fig.~\ref{fig:split_rc}, result from 
physical differences between stars rather than poor reddening corrections.
The DECam $u$-band spans approximately 3500-4000~\AA\ in wavelength, which is
a region that includes numerous strong metal-lines in red clump stars.  As a 
result, more metal-rich stars will have increased line blanketing and appear 
significantly redder in $u-i$ than their more metal-poor counterparts.

\begin{figure}
\includegraphics[width=\columnwidth]{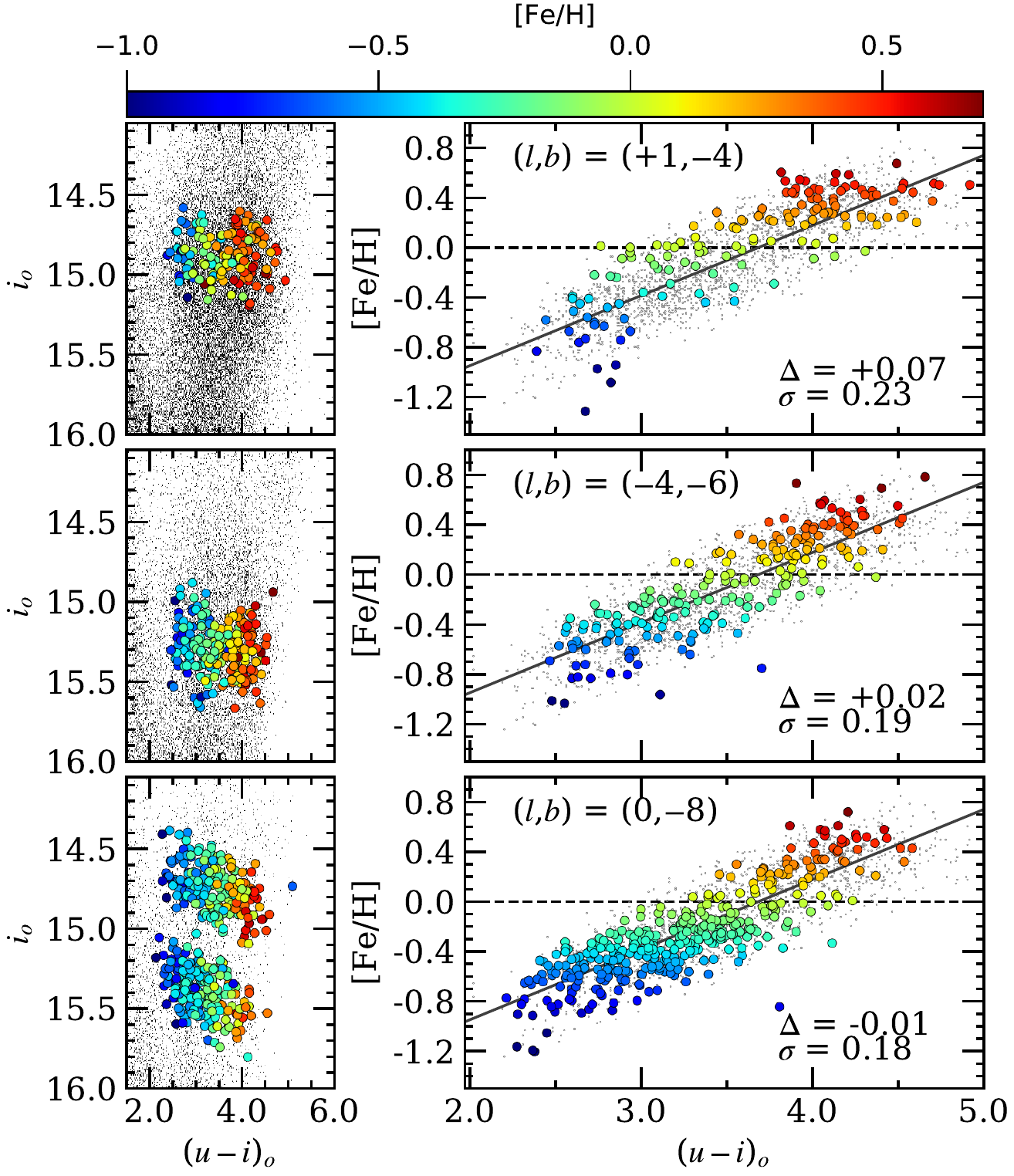}
\caption{\emph{Left:} dereddened $i$ versus $u-i$ CMDs centered in the
red clump region are shown for three bulge fields.  The large filled circles
are red clump stars from GIBS \citep{Zoccali17} that have spectroscopic [Fe/H]
determinations.  The colors indicate each star's [Fe/H] value and are saturated
at [Fe/H] = $-$1.0 (blue) and $+$0.7 (red).  Note the strong dependence of
($u-i$)$_{\rm o}$ on stellar metallicity.  \emph{Right:} plots of [Fe/H]
versus ($u-i$)$_{\rm o}$ are shown for the three fields.  The small grey
circles illustrate the trend of all 14 GIBS fields used here, and the best-fit
linear relation is shown with the solid dark grey line.  The mean offsets
($\Delta$) between the observed and predicted [Fe/H] values, along with the
standard deviations ($\sigma$), are provided for each field.}
\label{fig:ui_mdf}
\end{figure}

We can verify that the $u-i$ color dispersion is driven by metallicity via 
direct comparison with the GIBS database, which reported [Fe/H] values for
several thousand red clump stars in sight lines that largely overlap with 
BDBS.  Metallicity variations as a function of CMD location for three BDBS
fields are shown in Fig.~\ref{fig:ui_mdf} and indicate that ($u-i$)$_{\rm o}$
is strongly correlated with [Fe/H].  Additionally, Fig.~\ref{fig:ui_mdf}
shows that the color-metallicity relations are nearly identical across the
15 calibration sight lines, and are also independent of whether stars are in 
the bright or faint red clumps.  Therefore, we determined a global red clump 
color-metallicity relation for BDBS as:
\begin{equation}
[Fe/H] = 0.563(u-i)_{\rm o} - 2.074,
\end{equation}
where [Fe/H] is the calibrated iron abundance for a star and ($u-i$)$_{\rm o}$
is the dereddened BDBS color.  The typical scatter in [Fe/H] when using the 
derived color-metallicity relation is $\sim$ 0.2 dex or better, and is
probably limited by the accuracy of the GIBS metallicities, along with the
resolution and accuracy of the reddening map.  

\begin{figure}
\includegraphics[width=\columnwidth]{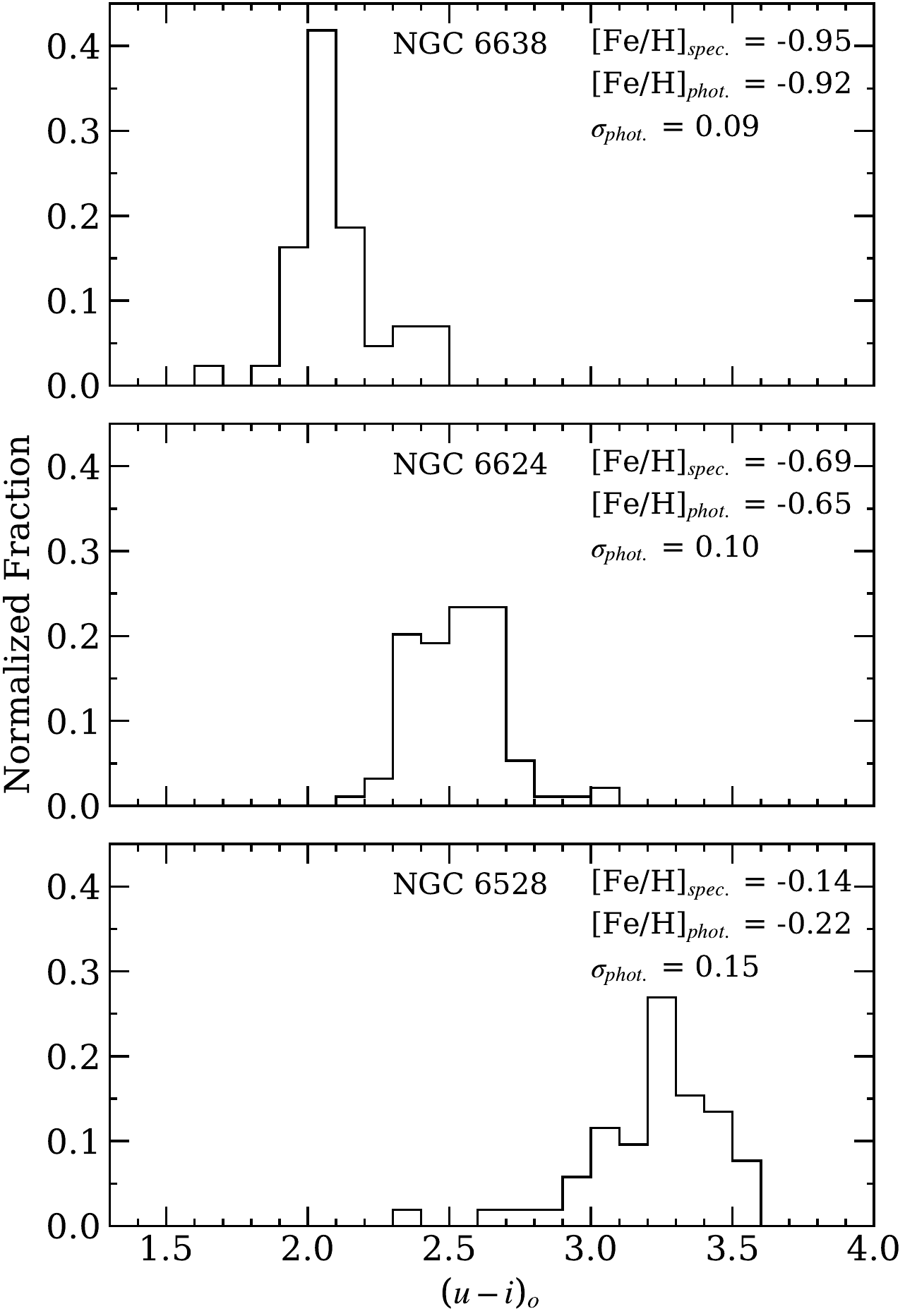}
\caption{Binned histograms illustrating the red clump $(u-i)_{o}$ distributions
are shown for three well-sampled BDBS globular clusters that span [Fe/H] $\sim$
$-$1 to $-$0.1.  Each histogram includes only stars that could be clearly
separated from the cluster RGB, had Gaia DR2 proper motions consistent with
cluster membership, and were bluer than the RR Lyrae region.  The
spectroscopic [Fe/H], photometric [Fe/H], and photometric [Fe/H] dispersion are
provided for each cluster.  Despite some increased color dispersion due to
factors such as mass loss, rotation rate, and light element abundance
variations, the $(u-i)_{o}$ distribution is strongly peaked for each cluster
and becomes redder at higher [Fe/H].  Spectroscopic metallicities for NGC 6638,
NGC 6624, and NGC 6528 are from \citet{Mauro14}, \citet{Valenti11}, and 
\citet{Munoz18}, respectively.}
\label{fig:gc_ui_conf}
\end{figure}

As an independent verification that $(u-i)_{o}$ is correlated with [Fe/H], 
Figure \ref{fig:gc_ui_conf} shows histograms of the $(u-i)_{o}$ distributions
for red clump stars in three bulge globular clusters.  The clusters NGC 6638, 
NGC 6624, and NGC 6528 were chosen because they span about a factor of 10 in
[Fe/H], have well-sampled red HB populations, reside within the BDBS reddening
map region, and have literature spectroscopic [Fe/H] measurements.  Using the 
color-metallicity relation provided in Equation 22, Figure \ref{fig:gc_ui_conf}
indicates that: the mean red clump color becomes redder at higher metallicity, 
the mean photometric [Fe/H] estimate is within 1$\sigma$ of the 
spectroscopic value, and the dispersion around the mean is approximately 
0.1-0.15 dex.  Therefore, we conclude that $(u-i)_{o}$ is an
accurate tracer of metallicity in old red clump stellar populations.

Although similar color-metallicity relations could be derived for brighter 
RGB stars, red clump stars are generally more useful because their absolute
magnitudes are relatively well constrained.  RGB-tip stars have been used in
the past to calculate metallicity distribution functions with both optical
\citep[e.g.,][]{Zoccali03,Johnson11} and near-IR \citep[e.g.,][]{Gonzalez13}
data.  However, since the [Fe/H] estimates using this isochrone method are 
functions of color \emph{and} magnitude, distance uncertainties can 
significantly affect a star's assumed metallicity.  Additionally, degeneracy 
in the color-metallicity relation can be large because the optical/near-IR
color dispersion is relatively small until the last few tenths of a magnitude
near the RGB-tip.  

Recently, \citet{Mohammed19} combined GALEX $NUV$ photometry
with Gaia $G$-band magnitudes and performed a similar color-metallicity
calibration for nearby red clump stars based on APOGEE abundances; however,
despite the $NUV-G$ color having a larger dispersion for red clump stars than
$u-i$, this method likely has limited utility in bulge fields.  For example, 
GALEX $NUV$ magnitudes have a much stronger dependence on extinction than the 
$u$-band, and such observations would have to reach at least $NUV$ $\sim$ 24-25
mag. with reasonable S/N to be useful in the bulge.  The limit would be fainter
in even moderately extincted fields.  GALEX pixels are also several times 
larger than those of DECam, which strongly affects photometry in crowded 
fields.  We conclude that $u-i$ is a superior color for photometrically 
measuring the metallicities of red clump stars in crowded bulge fields, and 
with a typical uncertainty of $\sim$ 0.2 dex our calibration is comparable in 
precision to spectroscopic methods such as the Calcium Triplet.

Although we have shown that $(u-i)_{o}$ correlates with stellar metallicity, we 
stress several important caveats related to photometric [Fe/H] measurements and
interpretations of the bulge's physical structure.  First, core He-burning 
stars with [Fe/H] $\la$ $-$1 tend to reside on the blue HB rather than the red 
clump, which means our analysis is biased against the most metal-poor stars.  
Fortunately, an overwhelming majority of stars in the bulge have $-$0.8 $\la$ 
[Fe/H] $\la$ $+$0.5.  Additionally, many stars with [Fe/H] $<$ $-$1 that are
projected onto the bulge are likely to be halo interlopers.

A second and more significant problem relates to the contamination rate and 
evolutionary state uncertainty of stars residing in the red clump region of 
bulge CMDs.  The most obvious contaminant is the exponential background of 
first ascent RGB, plus some AGB, stars that can overlap in color and magnitude 
with the red clump.  This simple background can account for as many as 20-30
percent of stars in the red clump region \citep[e.g.,][]{Nataf11,Clarke19}.
Since red clump stars are bluer than RGB stars of the same metallicity, the
RGB stars will follow a different color-metallicity relation.  
However, as extensively described by \citet{Nataf14}, quantifying the 
contaminating RGB population is difficult because metallicity, He abundance, 
and distance variations modify the colors, magnitudes, and evolutionary 
lifetimes of RGB, red clump, and AGB stars.  In the future, the red clump 
color-metallicity relation may be improved using a combination of 
asteroseismology, parallax distances, and proper motions to isolate samples of 
pure red clump stars residing at bulge distances.

\subsubsection{Metallicity Distribution Functions}

As mentioned in \S 1, the metallicity distribution function is an important 
tool for understanding the formation history of stellar populations.  
However, little agreement exists regarding the true nature of the bulge's 
metallicity distribution \citep[e.g., see review by][]{Barbuy18}.  For 
example, early work by \citet{Rich90} showed that the inner bulge was 
well-described as a unimodal population that largely followed the expected 
distribution from a one-zone closed box enrichment model.  In contrast, more 
recent measurements claim that the bulge is a multi-component system hosting 
anywhere from 2-5 distinct populations \citep[e.g.,][]{Hill11,Ness13_mdf,
Bensby17,Rojas17,Zoccali17,GarciaPerez18,Duong19}.  Most authors also find that
fields closer to the plane are dominated by metal-rich stars while those 
farther from the plane host more metal-poor stars.  Additional metal-poor 
populations with [Fe/H] $<$ $-$1 have also been detected 
\citep[e.g.,][]{GarciaPerez13,Ness13_mdf,Koch16}, but many of these stars are
likely halo interlopers.

\begin{figure*}
\includegraphics[width=\textwidth]{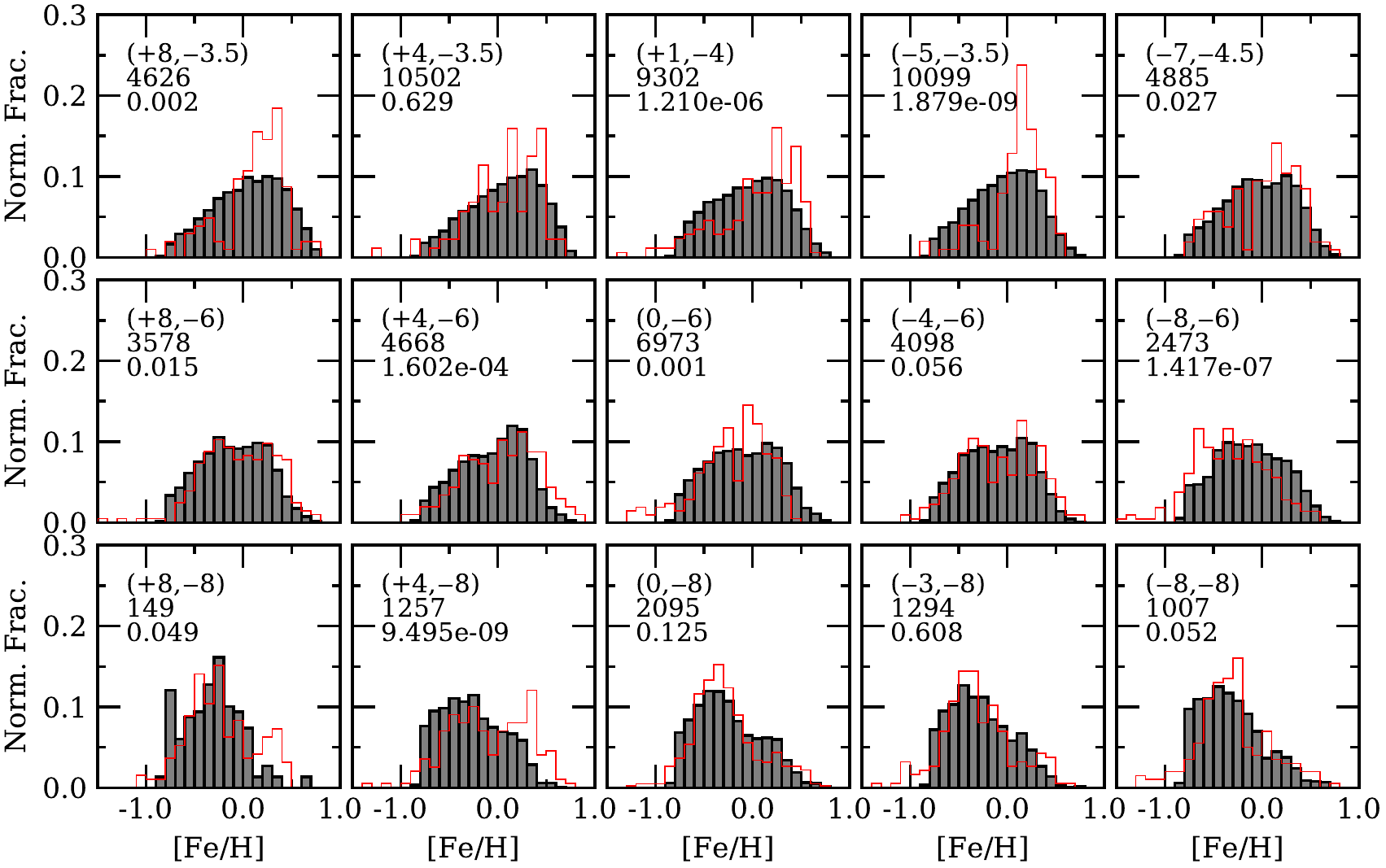}
\caption{Similar to Figure 4 in \citet{Zoccali17}, we compare the metallicity
distribution functions of red clump stars in BDBS (filled grey histograms),
derived using the color-[Fe/H] relation shown in Fig.~\ref{fig:ui_mdf}, with
those of GIBS (red histograms).  The Galactic coordinates, number of BDBS red
clump stars, and $p$-value of two-sided Kolmogorov-Smirnov tests are
provided in each panel.  Each BDBS field spans approximately 15$\arcmin$ in
radius.}
\label{fig:mdfs}
\end{figure*}

Fig.~\ref{fig:mdfs} summarizes the derived metallicity distribution functions
for the 15 calibration fields that total $>$ 67,000 red clump stars.  The 
distributions are largely compatible with previous analyses that found a 
vertical metallicity gradient, driven by the changing ratios of metal-poor and 
metal-rich stars, along with at least two ``peaks" in several, especially 
outer, bulge fields \citep[e.g.,][]{Zoccali08,Babusiaux10,Ness13_mdf,
Gonzalez15,Rojas17,Zoccali17}.  However, Fig.~\ref{fig:mdfs} indicates that 
fields interior to $b$ $\sim$ $-$6$\degr$, and reaching at least the $b$ $\sim$ 
$-$3.5$\degr$ limit shown here, do not necessarily possess a significant 
additional metal-poor component.

\begin{figure}
\includegraphics[width=\columnwidth]{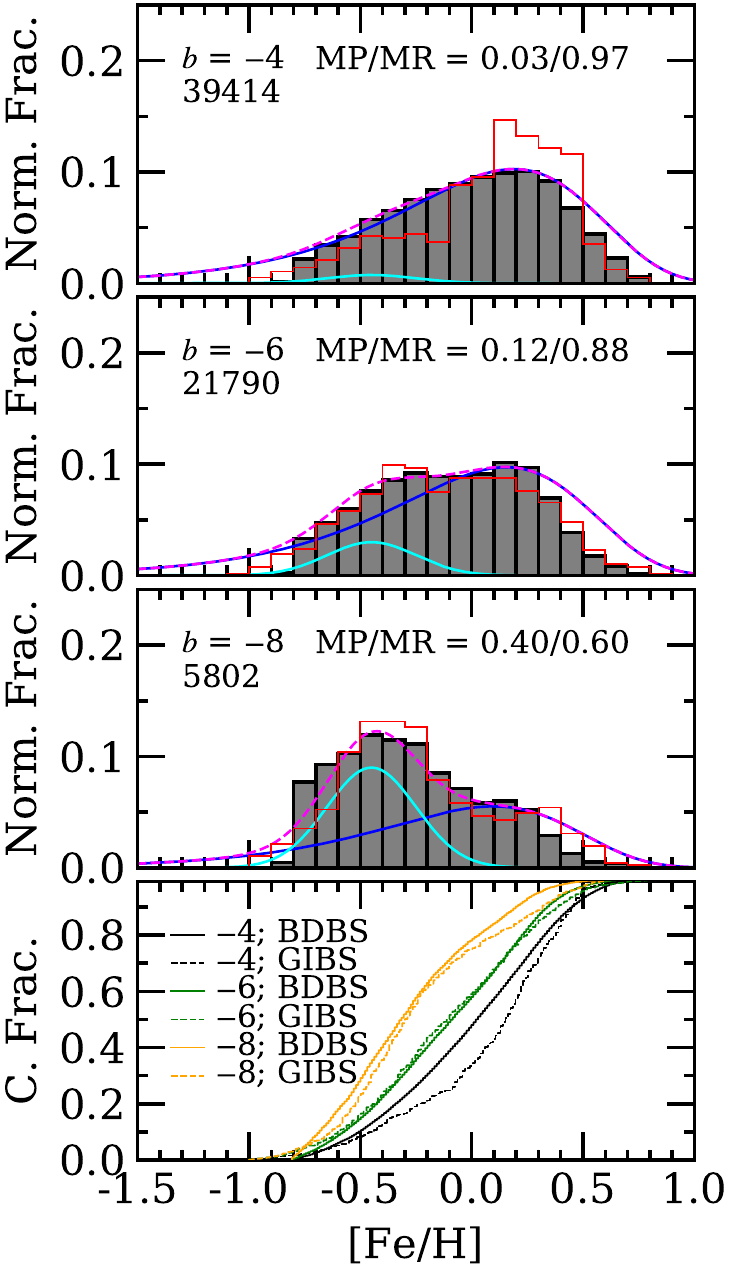}
\caption{Similar to Figure 7 in \citet{Zoccali17}, the [Fe/H] distributions
from Fig.~\ref{fig:mdfs} for BDBS (grey bars) and GIBS (red lines) have been 
summed across each Galactic latitude segment and are presented as histograms in
each panel.  The Galactic latitude and number of BDBS red clump stars are 
provided in each panel.  The solid blue lines show the expected distributions 
for a one-zone closed box gas exhaustion model with a mean effective yield of 
[Fe/H] $\sim$ $+$0.15.  The solid cyan lines illustrate the addition of a 
metal-poor, normally distributed component with $\langle$[Fe/H]$\rangle$ = 
$-$0.45 and $\sigma$ = 0.25, and the dashed magenta lines show the combined 
closed box and normal components.  The area ratios of metal-poor (cyan) and
metal-rich (blue) are provided in each panel.  The bottom panel compares the 
same data but as a cumulative distribution functions for the 
$b$ = $-$4$\degr$ (black), $b$ = $-$6$\degr$ (green), and $b$ = $-$8$\degr$ 
(orange) BDBS (solid lines) and GIBS (dashed lines) fields.}
\label{fig:mdf_all}
\end{figure}

Fig.~\ref{fig:mdf_all} sums across the fields at fixed Galactic latitudes and
shows that the $b$ = $-$4$\degr$ sight lines are morphologically similar to the
expected distribution from a simple closed box gas exhaustion model of the
form:
\begin{equation}
\frac{dN}{d[Fe/H]}\propto 10^{[Fe/H]}e^{-\frac{10^{[Fe/H]}}{y}},
\end{equation}
where $y$ is the mean effective yield \citep[e.g.,][]{Searle78,Rich90}.  
Distributions described by Equation 23 are unimodal but strongly asymmetric
with long metal-poor tails.  Similarities between closed box models and bulge
metallicity distributions have been noted before \citep[e.g.,][]{Rich90,
Zoccali03,Zoccali08}, but more recent investigations have dropped this model in 
favor of fitting multiple Gaussian functions.  Although such an approach is 
convenient for population studies and fitting, little physical motivation exists
regarding why the bulge should be a composite of two or more populations that 
each have a relatively narrow, normally distributed metallicity distribution 
function.  Systems composed of two or more populations with narrow, normally
distributed metallicity distributions do exist in nature, but such objects
often show extreme heavy element abundance variations that are indicative 
of a prolonged period of ``bursty" star formation and self-enrichment, 
particularly from low and intermediate mass AGB stars \citep[e.g., $\omega$ 
Cen;][]{Johnson10,Marino11_omcen}.  The Galactic bulge shows no evidence  
supporting this type of enrichment pattern \citep[e.g.,][]{Johnson12_bulge,
VanderSwaelmen16,Bensby17,Duong19}, although its physical parameters also 
differ substantially from those of ``iron-complex" globular clusters.

At least for $b$ = $-$4$\degr$, fitting Equation 23 to the BDBS data only 
leaves room for an additional metal-poor component at the few percent
level.  Although the simple closed box model follows the general shape of the
observed $b$ = $-$4$\degr$ distribution, we note some differences in the 
metal-poor and metal-rich tails.  For the metal-poor tail, BDBS does not find
as many stars as the model predicts.  However, the paucity of BDBS stars with
[Fe/H] $<$ $-$1 is likely driven by our selection function and the general
evolution of metal-poor stars to land on the blue HB.  At the metal-rich end,
Fig.~\ref{fig:mdf_all} also shows a steeper drop off than the model, but we 
suspect that this discrepancy is driven by limitations of the simple model.  A
truncated metal-rich tail could be accounted for by an improved model, such as
one that includes gas outflow.  Nevertheless, the cumulative distribution 
functions at $b$ = $-$4$\degr$ in Fig.~\ref{fig:mdf_all} for BDBS and GIBS do 
not exhibit well defined inflection points that would be indicative of strongly bimodal distributions.

For the fields at $b$ = $-$6$\degr$ and $-$8$\degr$, Fig.~\ref{fig:mdf_all}
does confirm previous claims that a secondary metal-poor population is found
near [Fe/H] $\sim$ $-$0.4.  However, following the assumption that the 
metal-rich peak is a tracer of the same underlying population that dominates
at $b$ = $-$4$\degr$, we find the secondary metal-poor component to be much 
weaker than previous estimates.  For example, \citet{Zoccali17} found that the 
metal-poor populations constitute 51 and 73 per cent of the stars at $b$ = 
$-$6$\degr$ and $-$8$\degr$, respectively.  In contrast, we find in those same 
fields that the secondary metal-poor component only contributes about 10 and 
40 per cent of the total star counts, respectively\footnote{Since the red clump
does not strongly sample stars with [Fe/H] $<$ $-$1, we do not have enough 
information to estimate the underlying shape of the separate metal-poor 
component.  Lacking compulsory evidence, we assume a Gaussian distribution.  
However, depending on the origin of the metal-poor component, it could have a 
different functional form.}.  We urge caution when force fitting multiple 
Gaussian functions to bulge metallicity distributions since 
Fig.~\ref{fig:mdf_all} highlights that a significant fraction of bulge stars
could follow an inherently skewed and non-normal distribution.

Sampling issues also remain an important consideration, and in this sense
further insight may be gained from the ($l$,$b$) = ($+$4,$-$3.5) panel in 
Fig.~\ref{fig:mdfs}.  For this field, \citet{Zoccali17} fit two Gaussian
distributions with mean [Fe/H] values of $\sim$ $-$0.2 and $+$0.3.  However,
the BDBS data show a clear peak near [Fe/H] $\sim$ $+$0.25 but do not show 
evidence of a local maximum at lower metallicity.  A two-sided 
Kolmogorov-Smirnov (KS) test returns a $p$-value\footnote{We adopt the common 
convention that a $p$-value $<$ 0.05 is sufficient to reject the null 
hypothesis.} of 0.629 when comparing the BDBS and GIBS [Fe/H] distributions, 
but the 120$\times$ larger sample in BDBS appears to smooth out other potential
modes.  Similarly, we note that the BDBS and GIBS distributions, as traced by
the KS-test $p$-values in Fig.~\ref{fig:mdfs}, may become more similar in the
lower density outer bulge regions.  Since the GIBS project observed $\sim$ 200
stars per field, regardless of the local stellar density, the metallicity
histograms presented in \citet{Zoccali17} sample progressively smaller 
fractions of the underlying distributions at lower latitudes.  Nevertheless, 
the BDBS fields also suffer from a variety of contamination issues (see 
$\S$4.3.2) that will require further detailed analyses to fully disentangle.

\subsection{Globular Cluster CMDs}

\begin{figure*}
\includegraphics[width=\textwidth]{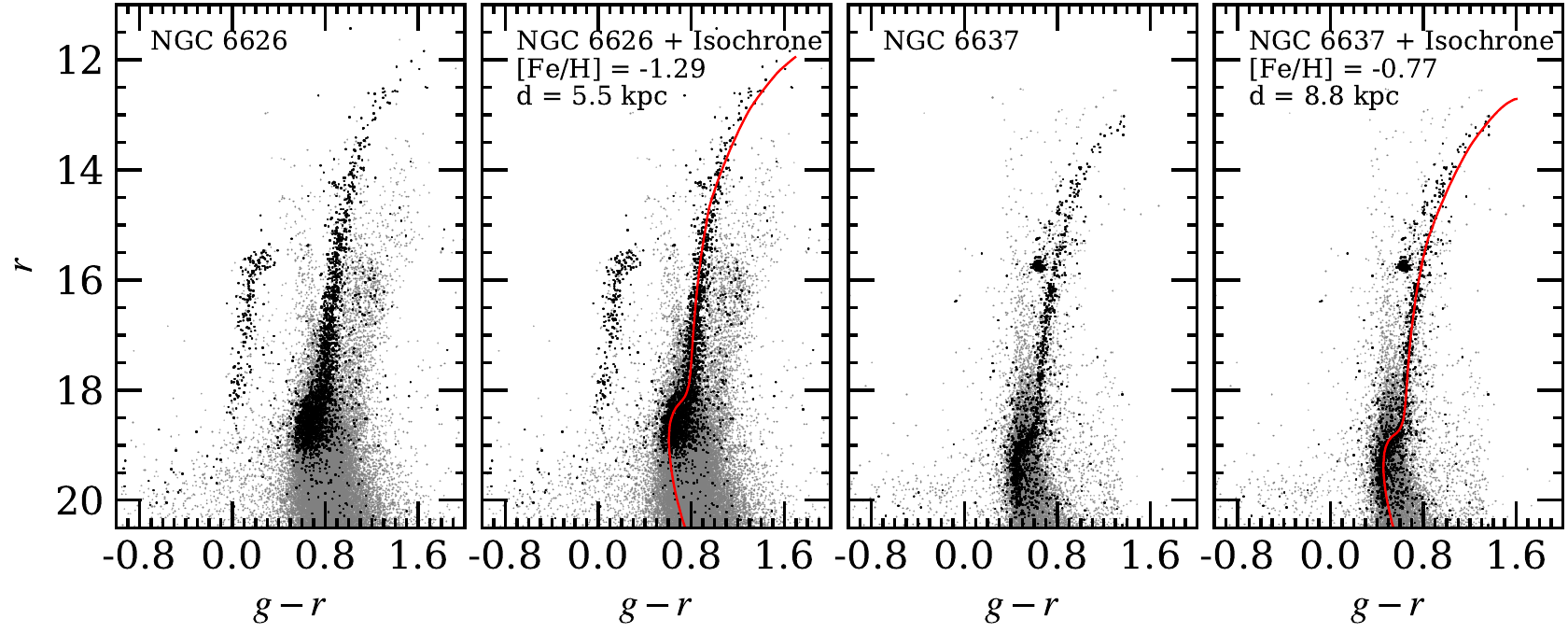}
\caption{Sample $r$ versus $g-r$ CMDs are shown for the bulge globular
clusters NGC 6626 (left two panels) and NGC 6637 (right two panels), which have
completely blue and red HBs, respectively.  The light grey symbols include all
stars within 5$\arcmin$ of the cluster centers and the black symbols indicate
stars that also have proper motions consistent with cluster membership.
Isochrones \citep{Girardi00} are overplotted (solid red lines) for both
clusters assuming ages of 12.5 Gyr and [$\alpha$/Fe] = $+$0.3, along with
the listed [Fe/H] and distance values from \citet{Villanova17} and
\citet{Lee6637}.  E(B-V) values of 0.40 and 0.18 mag. \citep{Harris96} were
applied to the NGC 6626 and NGC 6637 isochrones, respectively.}
\label{fig:gc_samp}
\end{figure*}

An additional area of inquiry for data validation and exploration with BDBS is 
globular cluster CMDs.  As shown in Fig.~\ref{fig:density_map}, BDBS includes
at least 25 Galactic globular clusters that span a variety of masses, 
metallicities, and HB morphologies.  For example, Fig.~\ref{fig:gc_samp} shows
$r$ versus $g-r$ CMDs for the globular clusters NGC 6626 ([Fe/H] $\sim$ $-$1.3)
and NGC 6637 ([Fe/H] $\sim$ $-$0.77), which have almost completely
blue and red HB morphologies, respectively.  For both clusters, BDBS easily
reaches below the main-sequence turn-off and also samples the RGB-tip.  
Additional key evolutionary indicators, such as the RGB-bump and AGB, are also 
readily visible in Fig.~\ref{fig:gc_samp}.  

A comparison with isochrones from \citet{Girardi00} indicates that the BDBS
globular cluster CMDs are well-calibrated onto the Pan-STARRS system, and
that the evolutionary sequences follow the expected distributions.  The small
star-to-star scatter along the RGB sequences also indicate that
the photometric precision is stable over a wide magnitude range.  Furthermore,
we do not detect any systematic variations in color or magnitude between
``bright" and ``faint" stars that would have been preferentially measured in
the long, short, or ultra-short exposures.

Fig.~\ref{fig:gc_samp} also shows the crowding and magnitude limitations of 
Gaia DR2 \citep{Gaia18_DR2} in the bulge region.  For NGC 6626, which has a 
Galactic latitude of $-$6$\degr$, Gaia DR2 proper motions are only useful for 
finding cluster members down to $r$ $\sim$ 19 mag.  However, in the less 
crowded NGC 6637 field at $b$ $\sim$ $-$10.3$\degr$ Gaia DR2 proper motions can
distinguish cluster members down to $r$ $\sim$ 20 mag.  In both cases BDBS
reaches at least 1-2 mag. deeper than Gaia.  Future Gaia data releases will
likely permit BDBS globular cluster investigations reaching down to the lower
main-sequence in most clusters.  We now use BDBS data to briefly investigate 
three additional globular clusters that have disputed properties.

\subsubsection{NGC 6656 (M 22)}

NGC 6656 is one of the most massive globular clusters in the Milky Way and is
thought to be the remnant core of a former dwarf spheroidal galaxy.  A key
indicator of the cluster's peculiarity is its metallicity distribution 
function, which shows a significant spread or bimodality that also coincides 
with variations in elements produced by the slow neutron-capture process
\citep{DaCosta09,Marino09,Lee16_M22}.  Combined near-UV and optical CMDs show
that NGC 6656 has a double sub-giant branch (SGB), and that stars on the two
branches have different heavy element abundances \citep{Piotto09,Marino12}.
However, \citet{Mucciarelli15_M22} claim that only elements produced by the 
slow neutron-capture process show variations in the cluster and that NGC 6656 
has a negligible [Fe/H] spread.

\begin{figure*}
\includegraphics[width=\textwidth]{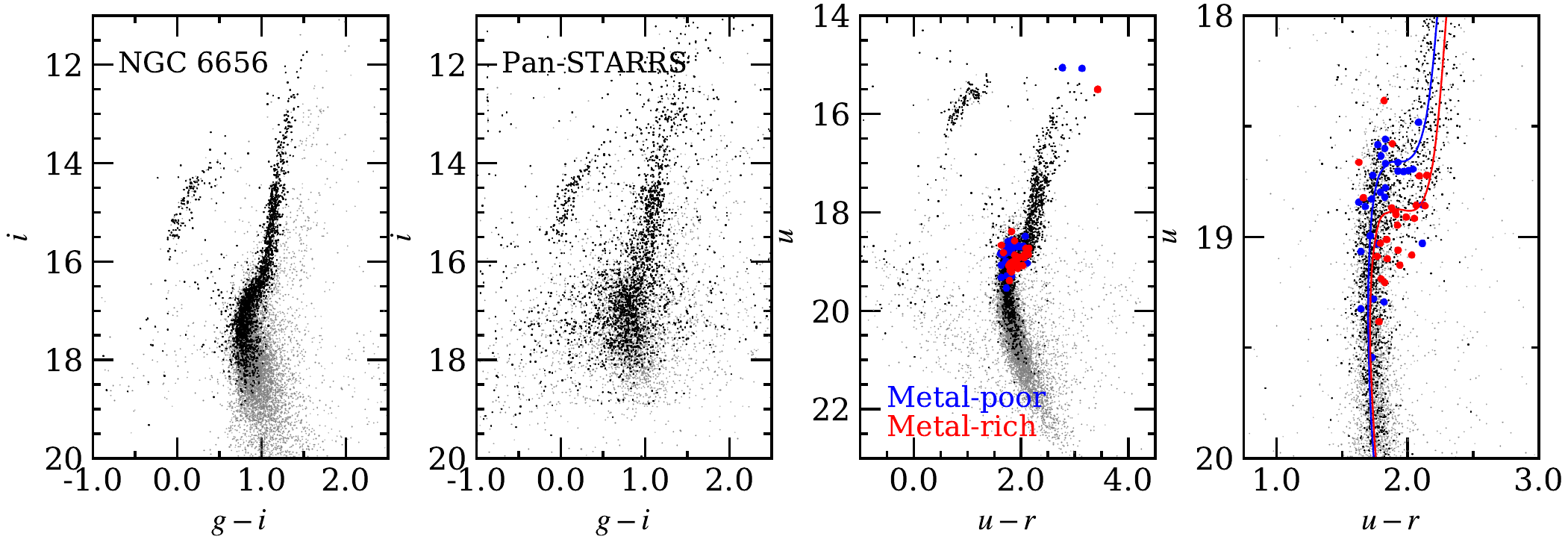}
\caption{Several BDBS CMDs are shown for stars within one half-light radius
(3.36$\arcmin$) of the globular cluster NGC 6656 (M 22).  The light grey
circles indicate all stars within a half-light radius while the black circles
show objects that also exhibit Gaia DR2 proper motions consistent with cluster
membership.  For comparison, the middle-left CMD includes stars with the same
spatial and proper motion selection criteria as the left panel, but only uses
data from Pan-STARRS.  The two right panels include metal-poor (s-poor)
and metal-rich (s-rich) stars from \citet{Marino09,Marino11,Marino12} as blue
and red circles, respectively.  The blue and red solid lines illustrate 12 Gyr,
$\alpha$-enhanced isochrones with [Fe/H] = $-$1.85 (blue) and [Fe/H] = $-$1.65
(red) from \citet{Dotter08}.  Note that the isochrones and split SGB are
consistent with a bimodal metallicity distribution for the cluster.}
\label{fig:n6656_cmd}
\end{figure*}

In Fig.~\ref{fig:n6656_cmd} we utilize our DECam $ugri$ photometry to 
investigate whether the split SGB is recovered in BDBS CMDs.  First considering
$i$ versus $g-i$, Fig.~\ref{fig:n6656_cmd} shows that a SGB color/luminosity 
spread is not easily detected using these filter combinations, despite both
BDBS and Gaia data providing a clean separation between the cluster and field 
stars well below the main-sequence turn-off.  Comparing with an equivalent
Pan-STARRS $i$ versus $g-i$ CMD for NGC 6656 reinforces that optical and 
near-IR colors are not optimal for detecting small metallicity variations in
the SGB region of old, metal-poor globular cluster populations.  However, the 
comparison does highlight the superior performance of BDBS compared to 
Pan-STARRS for bulge globular cluster CMD analyses as BDBS reaches at least
1-2 mag. deeper and has smaller photometric scatter.

When the $u$ and $r$-bands are combined, Fig.~\ref{fig:n6656_cmd} illustrates
that both the SGB and RGB split into two broadened sequences.  To verify that
the split SGB/RGB sequences are tracers of metallicity variations, we matched
BDBS to the spectroscopic results of \citet{Marino09,Marino11,Marino12}. 
The result shown in Fig.~\ref{fig:n6656_cmd} indicates that the 
split SGB/RGB using $u$ versus $u-r$ CMDs does follow the metallicity 
variations determined via spectroscopy.  The effect is particularly clear on 
the SGB where we find that stars in the ``metal-poor" NGC 6656 group 
overwhelmingly reside on the brighter SGB while those in the ``metal-rich" 
group trace the faint SGB.  A comparison with isochrones from \citet{Dotter08}
further reveals that the magnitude of the SGB split aligns with those predicted
from stellar evolution theory.  These results follow the findings of 
\citet[][see their Fig.~11]{Marino12} and confirms that the BDBS $u$-band data
are useful for measuring composition variations in both bulge red clump stars
and SGB/RGB stars in globular clusters.

\begin{figure*}
\includegraphics[width=\textwidth]{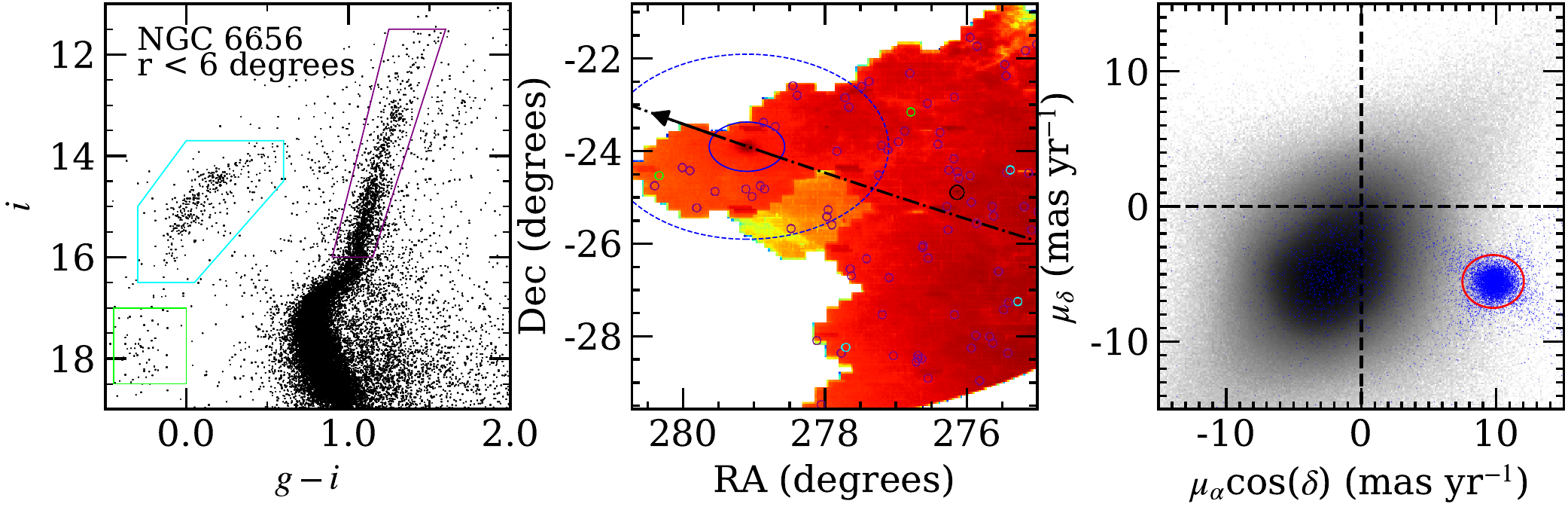}
\caption{\emph{Left:} An $i$ versus $g-i$ CMD for BDBS stars within 6
degrees of NGC 6656 that also exhibit proper motions consistent with cluster
membership.  The purple, cyan, and green regions outline the cluster's RGB,
blue HB, and extreme HB populations.  \emph{Middle:} A source density map for
the BDBS footprint within 6 degrees of NGC 6656.  The solid and dashed blue
lines indicate the King (31.9$\arcmin$) and Wilson (119.9$\arcmin$) tidal radii
from \citet{Kunder14}, respectively.  The open purple, cyan, and green circles
indicate possible RGB, blue HB, and extreme HB extra-tidal stars that lie
within the selection boxes in the left panel and have proper motions consistent
with cluster membership.  The large open black circle shows the position of
NGC 6626 (M 28).  The dot-dashed black line and arrow indicate the orbital
motion of NGC 6656 derived using Gaia DR2 proper motions.  Note that stars 
inside the King tidal radius are not plotted.  \emph{Right:} A Gaia-based 
vector point diagram for all BDBS stars within 6 degrees of NGC 6656 (shaded 
region) along with all stars within one half-light radius (blue circles).  The 
large red circle indicates the proper motion selection region for cluster 
membership.}
\label{fig:n6656_tidal}
\end{figure*}

Given the large footprint covered by BDBS, we can also investigate the possible
existence of extra-tidal cluster members.  For example, in 
Fig.~\ref{fig:n6656_tidal} we examine a region within 6 degrees of NGC 6656,
or at least out to the edges of our survey, to search for stars that have 
colors, magnitudes, and Gaia DR2 proper motions consistent with cluster 
membership.  Since the foreground disk and bulge RGB span a wide range of the 
CMD, we have restricted the search to only include stars that would be on the 
upper RGB or blue/extreme HB.  We also calculated the orbit of NGC 6656 using 
the Gala orbit integrator \citep{Price17} in order to aid the search for any 
tidal streams.  

\citet{Kunder14} used data from the Radial Velocity Experiment DR4 
\citep[RAVE;][]{Kordopatis13} and found only one possible extra-tidal star that 
had a color, magnitude, radial velocity, and chemical composition consistent 
with cluster membership.  Unfortunately, this star resides outside the BDBS
footprint and we are unable to verify whether it would have been detected using
the methods presented in Fig.~\ref{fig:n6656_tidal}.  However, we do find
numerous RGB and HB stars that have CMD locations and proper motions consistent
with cluster membership.  Fig.~\ref{fig:n6656_tidal} does not show any strong
evidence of a dense stream associated with NGC 6656, but we do note that many
extra-tidal candidates lie along the projected orbit.  A radial velocity and
chemical composition analysis of these stars would help confirm their status
as former cluster members.  Nevertheless, Fig.~\ref{fig:n6656_tidal} 
illustrates the utility of BDBS, especially when combined with Gaia, to 
search for extended stellar structures.

\subsubsection{NGC 6569}

NGC 6569 is a moderately metal-rich globular cluster ([Fe/H] $\sim$ $-$0.85) 
residing $\sim$ 3 kpc from the Galactic center, and is particularly interesting
because \citet{Mauro12} found evidence of a double red clump using near-IR VVV
data.  Such a discovery is important because the bulge cluster 
Terzan 5, which hosts at least 2-3 populations with different [Fe/H] values, 
was initially recognized as peculiar due to the detection of a double red clump
\citep{Ferraro09}.  However, \citet{Mauro12} noted that the double red clump 
feature was only prominent in near-IR CMDs and was not present in optical 
colors.  \citet{Johnson18_6569} obtained high resolution spectra for several 
hundred stars in NGC 6569, including those in both the bright and faint red 
clumps, but did not find any evidence supporting a metallicity spread nor a 
light element composition difference between the bright and faint red clump 
stars.  \citet{Johnson18_6569} concluded that the stars in both clumps were 
radial velocity members, but was not able to determine whether the near-IR 
double red clump feature was real.

\begin{figure*}
\includegraphics[width=100mm]{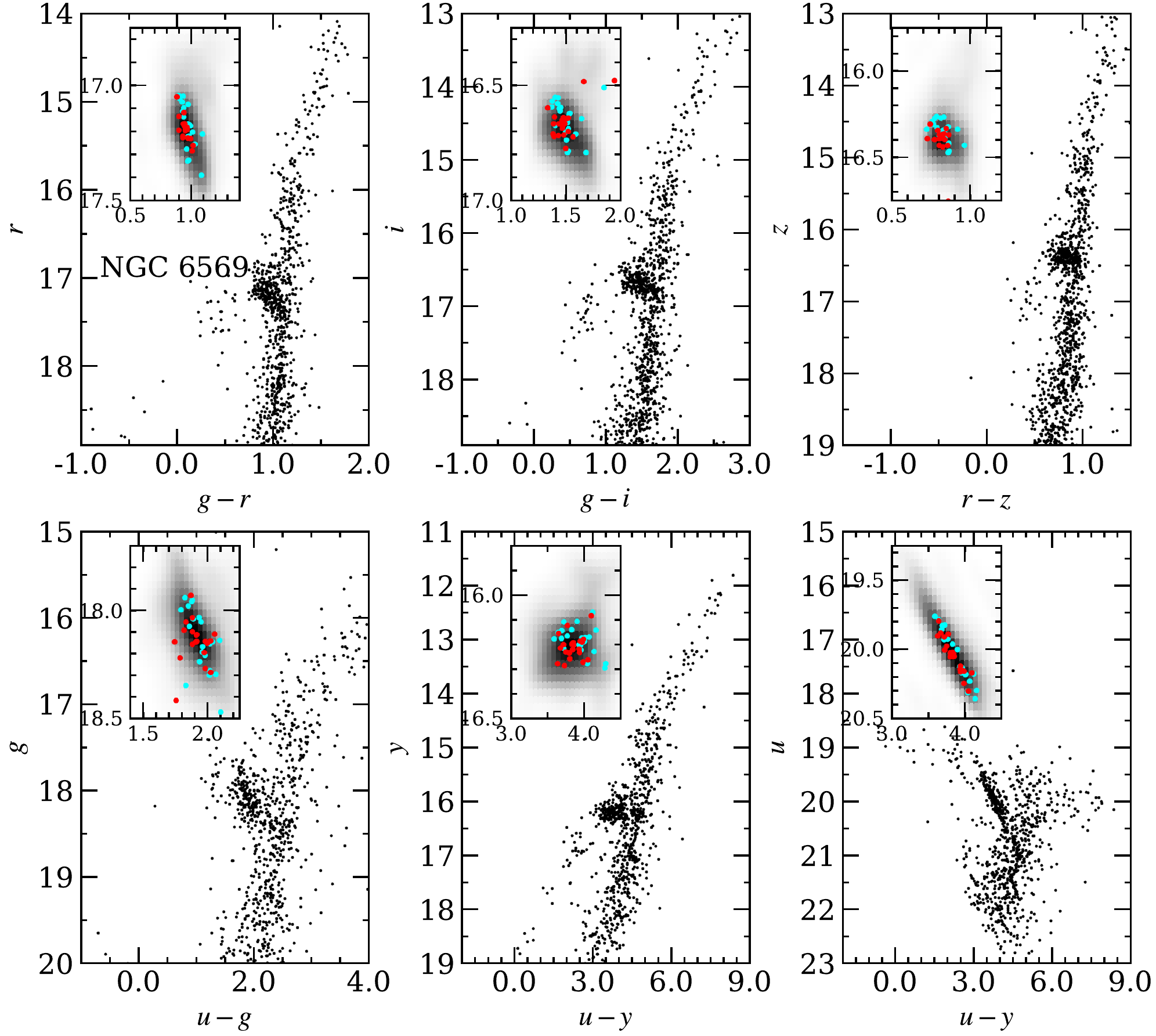}
\caption{Several BDBS CMDs using multiple filter combinations are shown for the
bulge globular cluster NGC 6569.  All CMDs include stars within 5$\arcmin$ of
the cluster center that also exhibit Gaia DR2 proper motions consistent with
cluster membership.  The inset figures for each panel are an expansion around
the red HB population.  The grey scale density map includes all stars within
the inset figure boundaries while the cyan and red open circles indicate
individual red HB stars that belong to the ``bright" (K$_{\rm S}$ $\sim$ 14.25)
and ``faint" (K$_{\rm S}$ $\sim$ 14.35) HB populations identified by
\citet{Mauro12}.  The BDBS CMDs do not show any particularly strong evidence
supporting a double red HB, with the exception of marginal detections in the 
$z$ and $y$-bands.}
\label{fig:n6569}
\end{figure*}

Therefore, in Fig.~\ref{fig:n6569} we utilize various color combinations in
all 6 filters to determine if any evidence exists to support the double red
clump claim by \citet{Mauro12}.  In order to identify stars belonging to the
bright (K$_{\rm S}$ $\sim$ 14.25 mag.) and faint (K$_{\rm S}$ $\sim$ 14.35 
mag.) HB populations from \citet{Mauro12}, we first generated a catalog of
BDBS stars within a radius of 5$\arcmin$ of the cluster center.  We then 
culled the catalog to retain only stars that had Gaia DR2 proper motions that 
were consistent with cluster membership.  Finally, this cleaned list was matched
with VVV so that BDBS stars residing within the bright and faint HB selection 
boxes of \citet{Mauro12} could be identified.  These potential double red clump
stars are shown as cyan (bright) and red (faint) circles within the inset 
CMDs of Fig.~\ref{fig:n6569}.

\begin{figure}
\includegraphics[width=\columnwidth]{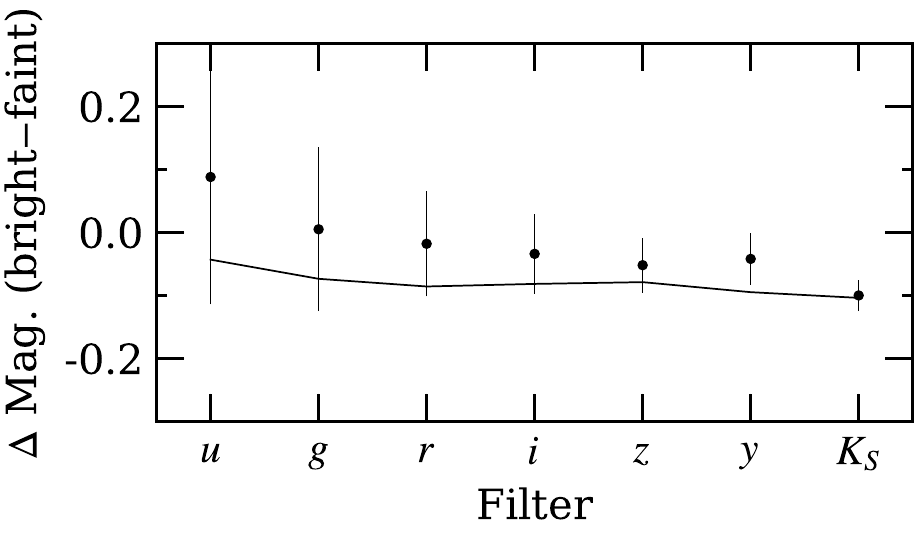}
\caption{Measured magnitude differences between NGC 6569 stars in the bright 
and faint red clumps are plotted for each BDBS filter ($ugrizy$) and for the VVV
K$_{\rm S}$-band.  The black dots represent the median differences while the 
error bars indicate the standard errors based on the dispersion of each
population.  The solid line indicates the expected differences for each filter
based on zero-age HB tracks from the Princeton-Goddard-PUC (PGPUC) isochrones
\citep{Valcarce12}.  The HB tracks were calculated assuming an RGB-tip mass
of 0.8 M$_{\rm \odot}$, $\Delta$Y = 0.02, [Fe/H] = $-$0.85, [$\alpha$/Fe] = 
$+$0.3, and a HB mass of 0.7 M$_{\rm \odot}$.}
\label{fig:n6569_split}
\end{figure}

Visual inspection of Fig.~\ref{fig:n6569} suggests that effectively no 
difference in brightness exists between the bright and faint red HB stars for
the $ugri$-bands.  This assertion is supported by the results of a Welch's
$t$-test comparing the magnitude distributions in each band, which returned 
$p$-values of 0.867, 0.658, 0.961, and 0.287 for the $ugri$-bands, 
respectively.  However, marginal detections of systematic differences in the
mean magnitudes of the $z$ and $y$-band distributions were found, and the 
Welch's $t$-tests returned $p$-values of 0.047 and 0.053, respectively.  

Fig.~\ref{fig:n6569_split} illustrates the median magnitude differences between
the bright/faint HB populations for each filter, including the VVV 
K$_{\rm S}$-band.  The data show a small but likely real increase in the 
magnitude differences between the bright and faint HB populations in 
progressively redder filters.  A comparison with theoretical zero-age HB tracks
from \citet{Valcarce12} indicates that the data are relatively consistent with 
an assumed helium spread of $\Delta$Y = 0.02, as suggested in 
\citet{Johnson18_6569}.  Therefore, we conclude that the red HB split detected
by \citet{Mauro12} is probably real, is marginally detected in the 
BDBS $z$ and $y$-band data, and that the diminishing magnitude differences in
bluer wavelengths can possibly be explained by a small but discrete He 
abundance spread in the cluster.

\subsubsection{FSR 1758}

FSR 1758 was originally detected as a probable star cluster in a low latitude
2MASS search by \citet{Froebrich07}, but has only recently received significant
attention as a possible galaxy nucleus or massive globular cluster 
\citep{Cantat18,Barba19,Simpson19,Villanova19}.  \citet{Barba19} claimed
to detect a large number of extra-tidal stars, and suggested that FSR 1758 may
even be part of a larger extended structure, such as an undiscovered dwarf 
galaxy.  On the other hand, \citet{Simpson19} noted that many of these 
extra-tidal stars are likely foreground dwarfs rather than cluster giants, and
\citet{Villanova19} performed a radial velocity and chemical composition 
analysis that found FSR 1758 to likely be a monometallic globular cluster.  
Since BDBS covers a large fraction of the region around FSR 1758, we can 
provide some additional insight into the presence (or not) of extra-tidal 
debris.

\begin{figure*}
\includegraphics[width=\textwidth]{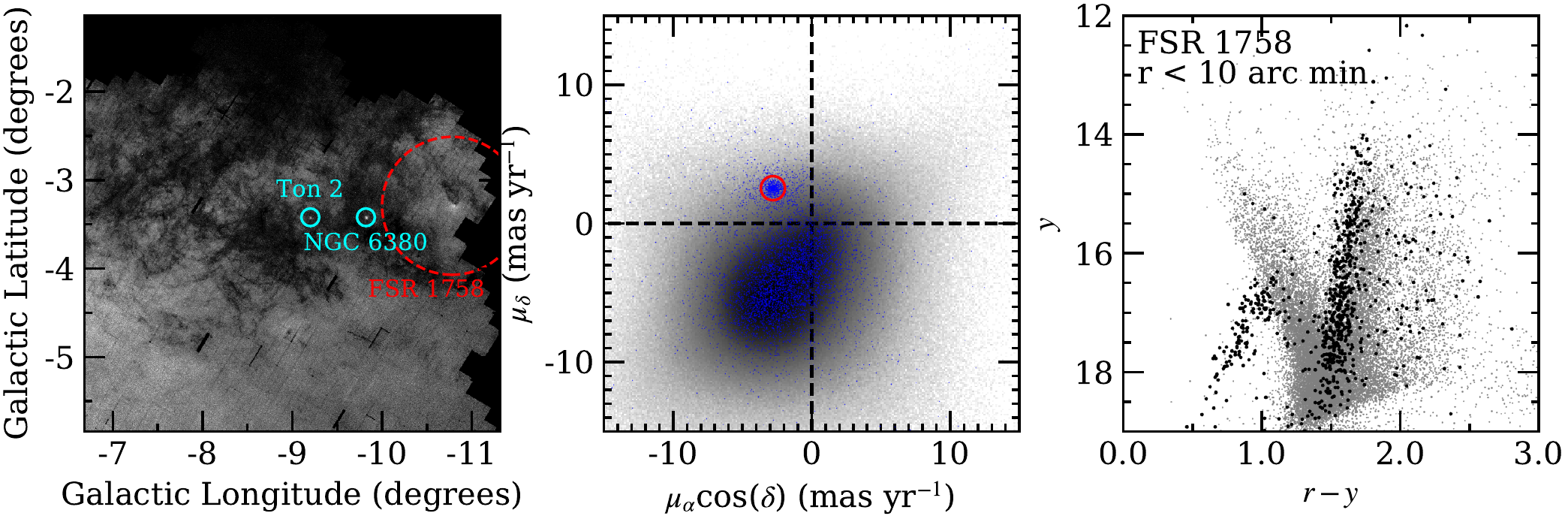}
\caption{\emph{Left:} a flux-weighted $r$-band density map of the BDBS region
within $\sim$ 5 degrees of the massive globular cluster FSR 1758.  The 0.78
degree tidal radius suggested by \citet{Barba19} is illustrated by the dashed
red line.  The nearby but unrelated globular clusters NGC 6380 and Ton 2 are
also indicated by the open cyan circles. \emph{Middle:} a Gaia DR2 vector point
diagram is shown as a shaded two dimensional histogram for all objects with
proper motion measurements in the left panel.  The blue symbols highlight stars
with radial distances $<$ 10$\arcmin$ from the cluster center, and the open red
circle indicates the selection criteria used to identify cluster members.
\emph{Right:} a $y$ versus $r-y$ CMD is shown for BDBS targets
within 10$\arcmin$ of the cluster center (grey symbols) and for which Gaia DR2
proper motions are consistent with cluster membership (black symbols).  Note
the extended blue HB and gap near $y$ $\sim$ 17 that is reminiscent of
NGC 6656 (e.g., see Fig.~\ref{fig:n6656_tidal}).}
\label{fig:fsr1758_1}
\end{figure*}

Although FSR 1758 is near the edge of the BDBS footprint, 
Fig.~\ref{fig:fsr1758_1} shows that the object is easily detected in our 
survey.  Furthermore, members are readily identified through a combination of 
broad-band photometry and especially Gaia DR2 proper motions.  For this paper, 
we restrict our CMD analysis to the $r$-band and redder filters since FSR 1758
lies outside our reddening map.  Redder colors are also generally
preferred when searching for extra-tidal members because 
Fig.~\ref{fig:fsr1758_1} shows that the region near FSR 1758 suffers from
significant and highly variable differential reddening.  

The $r$ versus $r-y$ CMD shown in Fig.~\ref{fig:fsr1758_1} indicates that the
FSR 1758 RGB is significantly bluer than a majority of the bulge RGB stars,
which is consistent with past work suggesting [Fe/H] $\sim$ $-$1.5 
\citep{Barba19,Villanova19}.  Additionally, we confirm that FSR 1758 has an
extended and very blue HB, along with a possible gap in the HB distribution
near $y$ $\sim$ 17 mag.  In this sense, the HB morphology of FSR 1758 closely
resembles that of NGC 6656 (e.g., see Fig.~\ref{fig:n6656_tidal}), including 
the nearly complete absence of red HB stars.  Although differential reddening
smears out the RGB color distribution shown in Fig.~\ref{fig:fsr1758_1}, the
data do not provide any clear evidence supporting a large metallicity spread.

\begin{figure}
\includegraphics[width=\columnwidth]{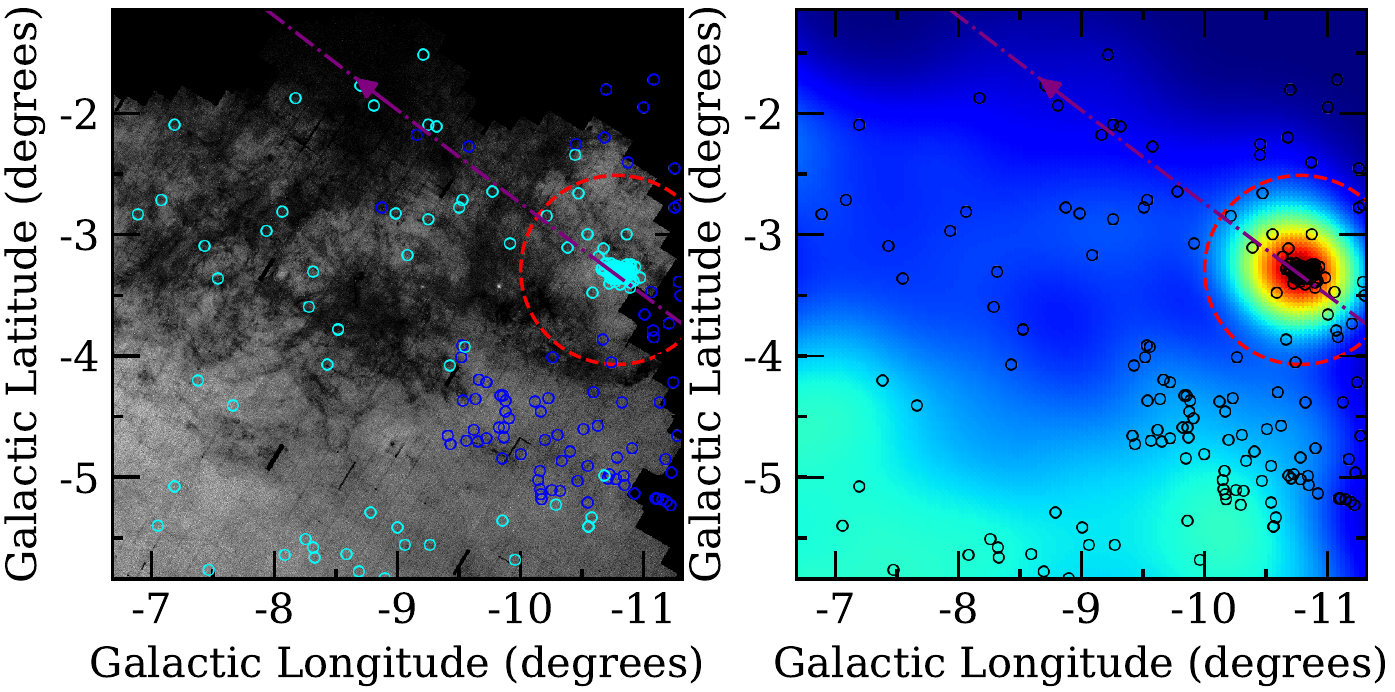}
\caption{\emph{Left:} a similar flux-weighted r-band density map as shown in
Fig.~\ref{fig:fsr1758_1} is provided for FSR 1758.  The open blue circles
indicate possible RGB stars with colors and proper motions consistent with
cluster membership, as identified by \citet{Barba19}.  Similarly, the open cyan
circles show the locations of blue HB stars identified in the BDBS catalog that
have r-y colors and proper motions consistent with cluster membership.  The
dot-dashed purple line and arrow shows the integrated orbit for FSR 1758.
\emph{Right:} a color gradient map is shown that illustrates the density of all
Gaia-BDBS sources with proper motions residing in the cluster selection region
of Fig.~\ref{fig:fsr1758_1}, regardless of a source's color or luminosity.  
Bluer shaded colors indicate lower source densities.  The open black circles 
illustrate the positions of all potential RGB and blue HB stars shown in the 
left panel.  Note that a number of blue HB stars lie along the orbit path.}
\label{fig:fsr1758_2}
\end{figure}

In Fig.~\ref{fig:fsr1758_2} we investigate the spatial distribution of stars 
that may be associated with FSR 1758, especially those outside the tidal 
radius.  Although Fig.~\ref{fig:fsr1758_2} shows that FSR 1758's RGB is 
generally bluer than the bulge sequence, it overlaps enough that cluster RGB
stars cannot be readily identified based on their CMD location alone.  However,
the blue HB stars are well-separated from both the bulge RGB and foreground
disk sequences, and are therefore more useful tracers outside the tidal radius.
We also utilize Gaia DR2 proper motions to remove any remaining bulge blue HB 
stars that happen to have similar colors/magnitudes to the FSR 1758 HB stars.

The resulting stars with colors, magnitudes, and proper motions consistent with
membership in FSR 1758 are identified as cyan open circles in 
Fig.~\ref{fig:fsr1758_2}.  An obvious clump is found within a few arc minutes
of the cluster center, but the sharp drop-off in blue HB density with radius
may indicate that the tidal radius is significantly smaller than the 
0.78$\degr$ reported by \citet{Barba19}.  We do find potential extra-tidal 
stars up to several degrees away, but unlike the \citet{Barba19} results our
potential extra-tidal stars are distributed throughout the field rather than
concentrated in a conspicuous clump.

When we plot the coordinates of all potential extra-tidal stars over a map 
showing the density of all Gaia DR2 detections that have proper motions 
matching the FSR 1758 selection region (right panel of 
Fig.~\ref{fig:fsr1758_2}), we find that an overwhelming number of the 
extra-tidal stars lie in regions where the background contamination is high.  
A comparison of the two panels in Fig.~\ref{fig:fsr1758_2} suggests that the 
``low-density" background regions are generally areas with high extinction.  
Therefore, we find in agreement with \citet{Simpson19} that most or all of the 
extra-tidal stars identified in \citet{Barba19} are likely false positives.  We
do not find any evidence that FSR 1758 is part of a larger structure.  However,
we do find several possible extra-tidal members that lie along the leading edge
of the cluster's orbit so significant tidal debris may still exist.

\section{Summary}

BDBS provides $ugrizy$ photometry, calibrated onto the SDSS ($u$) and 
Pan-STARRS ($grizy$) systems, for $\sim$ 250 million stars spanning more
than 200 square degrees of the Southern Galactic bulge down to depths of 
$\sim$ 22-23 mag.  For most fields, the data have already been corrected for
reddening using the extinction maps from \citet{Simion17}.  This paper 
describes the tools and methods developed to collate $>$ 10 billion detections 
in $\sim$ 450,000 CCD images into a usable catalog, and also includes some of 
the first science results from the project.  One of the most important 
discoveries is the tight correlation ($\sigma$ $\sim$ 0.2 dex) between 
dereddened $u-i$ color and [Fe/H] for bulge red clump giants.  In the future,
the derived color-metallicity relation will be applied to millions of red clump
stars, and should prove to be a transformational technique for deriving 
accurate [Fe/H] values from BDBS data.  

For this paper, we applied the red clump color-metallicity relation to 15 
sight lines and found that the bulge is not uniformly bimodal.  Fields near
$b$ $\sim$ $-$4$\degr$ are well fit by a simple closed box enrichment models
that possess a metal-rich peak and a long metal-poor tail.  Any additional
metal-poor components that might exist are limited to the few per cent level.
Exterior bulge fields appear bimodal, and require the addition of a secondary 
metal-poor population in order to adequately fit their metallicity 
distributions.  However, assuming that the metal-rich peaks in these fields 
are scaled versions of the $b$ $\sim$ $-$4$\degr$ distributions, the long 
metal-poor tails inherent to the closed box model significantly decreases the
strength of the required secondary populations compared to pure Gaussian
mixture models.

The BDBS data were further validated using investigations of globular clusters
residing in the survey area.  For example, we showed that NGC 6626 (blue HB)
and NGC 6637 (red HB) are well-fit by isochrones, and that our observations
easily reach from the RGB-tip to below the main-sequence turn-off.  Standard
photometric sequences, such as the AGB and RGB-bump, are also clearly seen.
Investigations of NGC 6656 (M 22) also showed that we detect the SGB and RGB
splits when using the $u$-band, and that the various sequences are driven by
metallicity differences.  We also searched for extra-tidal stars and found 
several candidates up to several degrees away that lie along the orbit path.
An analysis of NGC 6569 showed that the split red HB detected in the near-IR by
\citet{Mauro12} is probably real, and that we find marginal detections using
the $z$ and $y$-bands.  However, this split HB was restricted to redder colors,
which is consistent with a small He abundance spread ($\Delta$Y $\sim$ 0.02).
Finally, we used BDBS data to investigate FSR 1758, and found in agreement with
past work that the cluster is metal-poor relative to the bulge, has a very 
blue and extended HB, and is likely a monometallic globular cluster.  We find 
some evidence of extra-tidal stars lying along the leading path of the 
cluster's orbit, but do not confirm the existence of a significant tidal
debris field.  Instead, most of the tidal debris found by \citet{Barba19} seems
likely to be foreground field stars.

The full BDBS data set will be publicly released soon (Johnson et al., in 
prep.), and will be a critical community data set for exploring stellar 
populations within the inner disk and Galactic bulge.  The BDBS catalog will
be especially useful when combined with kinematic and imaging surveys
operating in other wavelengths.  The work presented here offers insight into
the types of problems that can be addressed using BDBS, which also serves as a 
pathfinder for observing strategies and science goals that may be achieved 
with the Vera C. Rubin Observatory.

\clearpage
\section*{Acknowledgements}
C.I.J. gratefully acknowledges support from the Clay Fellowship, administered 
by the Smithsonian Astrophysical Observatory, and thanks Nelson Caldwell for
numerous helpful discussions.  C.I.J. and R.M.R. thank Rodrigo Ibata for 
helpful discussions and early work on the project.  C.I.J., R.M.R., W.I.C., 
M.D.Y., S.M., and C.A.P. acknowledge support by the National Science Foundation
(NSF, grant AST-1412673).  C.A.P. acknowledges the generosity of the Kirkwood 
Research Fund at Indiana University.  A.K. gratefully acknowledges funding by 
the Deutsche Forschungsgemeinschaft (DFG, German Research Foundation) -- 
Project-ID 138713538 -- SFB 881 (``The Milky Way System''), subprojects A03, 
A05, A11.  This research was supported in part by Lilly Endowment, Inc., 
through its support for the Indiana
University Pervasive Technology Institute, and in part by the Indiana METACyt 
Initiative. The Indiana METACyt Initiative at IU was also supported in part by 
Lilly Endowment, Inc.  This material is based upon work supported by the 
National Science Foundation under Grant No. CNS-0521433.  This work was 
supported in part by Shared University Research grants from IBM, Inc., to 
Indiana University.  This project used data obtained with the Dark Energy 
Camera (DECam), which was constructed by the Dark Energy Survey (DES) 
collaboration.  Funding for the DES Projects has been provided by the U.S. 
Department of Energy, the U.S. National Science Foundation, 
the Ministry of Science and Education of Spain, the Science and Technology 
Facilities Council of the United Kingdom, the Higher Education Funding Council 
for England, the National Center for Supercomputing Applications at the 
University of Illinois at Urbana-Champaign, the Kavli Institute of Cosmological
Physics at the University of Chicago, the Center for Cosmology and 
Astro-Particle Physics at the Ohio State University, the Mitchell Institute for
Fundamental Physics and Astronomy at Texas A\&M University, Financiadora de 
Estudos e Projetos, Funda{\c c}{\~a}o Carlos Chagas Filho de Amparo {\`a} 
Pesquisa do Estado do Rio de Janeiro, Conselho Nacional de Desenvolvimento 
Cient{\'i}fico e Tecnol{\'o}gico and the Minist{\'e}rio da Ci{\^e}ncia, 
Tecnologia e Inovac{\~a}o, the Deutsche Forschungsgemeinschaft, 
and the Collaborating Institutions in the Dark Energy Survey.  The 
Collaborating Institutions are Argonne National Laboratory, 
the University of California at Santa Cruz, the University of Cambridge, 
Centro de Investigaciones En{\'e}rgeticas, Medioambientales y 
Tecnol{\'o}gicas-Madrid, the University of Chicago, University College London, 
the DES-Brazil Consortium, the University of Edinburgh, 
the Eidgen{\"o}ssische Technische Hoch\-schule (ETH) Z{\"u}rich, 
Fermi National Accelerator Laboratory, the University of Illinois at 
Urbana-Champaign, the Institut de Ci{\`e}ncies de l'Espai (IEEC/CSIC), 
the Institut de F{\'i}sica d'Altes Energies, Lawrence Berkeley National 
Laboratory, the Ludwig-Maximilians Universit{\"a}t M{\"u}nchen and the 
associated Excellence Cluster Universe, the University of Michigan, 
{the} National Optical Astronomy Observatory, the University of Nottingham, 
the Ohio State University, 
the OzDES Membership Consortium
the University of Pennsylvania, 
the University of Portsmouth, 
SLAC National Accelerator Laboratory, 
Stanford University, 
the University of Sussex, 
and Texas A\&M University.  Based on observations at Cerro Tololo 
Inter-American Observatory, National Optical Astronomy Observatory 
(2013A-0529;2014A-0480; R.M. Rich), which is operated by the Association of 
Universities for Research in Astronomy (AURA) under a cooperative agreement 
with the National Science Foundation.\\

Data Availability: The raw and pipeline reduced DECam images are available for 
download on the NOAO archive at http://archive1.dm.noao.edu/.  Astrometric,
photometric, and reddening catalogs are in the process of being prepared for
public release.  However, early release may be provided upon request to the
corresponding author.




\bibliographystyle{mnras}
\bibliography{references}


%
%


\bsp    
\label{lastpage}
\end{document}